\documentclass[journal=jctcce,manuscript=article]{achemso}
\setkeys{acs}{articletitle=true, doi=true, maxauthors=15, etalmode=truncate}

\usepackage{amsmath}
\usepackage{graphicx}
\usepackage{gensymb}
\usepackage{float}
\usepackage{color}
\usepackage[flushleft]{threeparttable}
\usepackage{etoolbox}   
\usepackage{booktabs}
\usepackage{chemist}
\usepackage{derivative}  
\usepackage{amsfonts}
\usepackage{siunitx}
\usepackage[shortlabels]{enumitem}
\usepackage{algorithm}
\usepackage{algpseudocode}
\usepackage[version=4]{mhchem}

\newcommand{\probP}{\text{I\kern-0.15em P}}

\usepackage{ragged2e}
\usepackage{indentfirst}
\usepackage[all]{nowidow}

\usepackage{upgreek}   
\renewcommand{\Delta}{\Updelta}
\renewcommand{\Omega}{\Upomega}
\renewcommand{\Gamma}{\Upgamma}
\renewcommand{\Lambda}{\Uplambda}
\renewcommand{\Xi}{\Upxi}

\newcommand{\norm}[1]{\left\lVert#1\right\rVert}

\usepackage[linktocpage,colorlinks=true,linkcolor=blue,urlcolor=blue,citecolor=blue,hyperfigures]{hyperref}
\usepackage{multirow}

\title{Accelerating Chemical Potential Calculations with Minimal Normalizing Flows}
\author{Philippe B. Baron}
\affiliation{Department of Chemical and Biological Engineering, Princeton University, Princeton, NJ, 08540}

\author{Athanassios Z. Panagiotopoulos}
\email{azp@princeton.edu}
\affiliation{Department of Chemical and Biological Engineering, Princeton University, Princeton, NJ, 08540}

\date{\today}
\begin{document}
\begin{abstract}
Chemical potentials are among the most important properties that can be obtained from a molecular simulation since they define many technologically relevant collective properties such as solubilities and activity coefficients. The chemical potential of a species in solution is typically obtained by computing the free energy change of adding that species into a bulk system, a calculation typically very expensive for systems of technological interest such as electrolytes, due to the lack of phase space overlap between ``not-inserted" and ``inserted" states. Recently, normalizing flows have been introduced as a way to accelerate free energy computations by learning a bijective function to map the configuration space of one Boltzmann distribution onto another. Currently, these trainable mappings are constructed to be as expressive as possible, theoretically having the ability to create a perfect mapping between states. This expressivity makes them difficult to train, limiting their ability to be generated ``on-the-fly" for any new free energy estimation challenge, and in practice these mappings have shown only modest performance improvements for liquid systems. We address these issues by introducing the concept of a ``minimal" normalizing flow (MNF). This is a trainable bijective mapping that is intentionally limited in expressivity, and instead applies low-dimensional, physically-informed transformations. Useful MNFs can be trained in approximately 1 minute of GPU time due to their simplicity and our introduction of a novel loss function based on the Bhattacharyya distance as an alternative to the Kullback-Leibler Divergence. We show how calculations of chemical potentials of  pure and binary Lennard-Jones particle systems can be accelerated by at least an order of magnitude with a simple radial mapping. For a more complex illustration of the approach, we apply a two-dimensional (radial and orientational) mapping to the solvation of sodium and chloride ions in water, showing that MNFs can increase the effective sample size by as much as 3 times for charging free energy calculations and as much as 8 times for calculating free energy changes due to force field perturbations. This provides the foundation for the development of physically-informed normalizing flows that can accelerate complex free energy calculations while retaining low training costs. 
\end{abstract}

\maketitle

\section{Introduction}
\label{intro}

The free energy is a quantity that determines the stability of a microscopic system, thus making its accurate and efficient theoretical prediction important to the computational design of new drugs and materials \cite{RN201,RN197,RN200}. Given a suitable force field (FF) for a system, Monte Carlo (MC) and Molecular Dynamics (MD) simulations are powerful tools that can provide free energy estimates for a wide variety of thermodynamic conditions by generating independent samples from equilibrium Boltzmann distributions. Free energy calculations from simulations provide access to rich thermodynamic information such as the phase diagrams of complex, multi-component systems, connecting microscopic interactions to macroscopic behavior relevant to technological applications \cite{RN202,RN198}.

Force field functions typically have empirically tunable parameters that need to be adjusted to obtain thermodynamic property predictions that align with experimental data. The chemical potential is defined as the free energy change for inserting an additional molecule of a species into an existing system. Computing this property has been established as the computational bottleneck in obtaining phase diagrams for complex, technologically relevant systems such as electrolyte solutions \cite{RN166,RN209}. In addition, since this property determines solution phase behavior, and other key properties such as activity coefficients, it has been established as one of the most stringent tests of FF performance \cite{RN6,RN7,RN122,RN159}.

These considerations suggest that it is important to be able to \emph{accurately and efficiently} estimate the free energy difference between two states (which we label states 0 and 1) from simulation. 
This problem was first approached by Zwanzig in 1954, when he introduced what is now commonly referred to as the Free Energy Perturbation (FEP) estimator \cite{RN110}. This is an importance sampling estimate of the free energy difference when we only have access to samples from one of the states. If all samples from state 0 are very unlikely to be seen in state 1, the importance weights of these samples are negligible, which makes the FEP estimator susceptible to high variance and bias. Alternatively, if we are able to obtain samples from both states in question, the Bennett Acceptance Ratio (BAR) provides the provable lowest-variance estimate of the free energy difference \cite{RN18}. The Multi-state Bennett Acceptance Ratio (MBAR) extends this approach when we want to obtain the mutual free energy differences between $M$ states \cite{RN45}. These methods are the standard for the estimation of free energy differences from a set of discrete states, but as we discuss later, the error in these estimates is inversely proportional to the configuration space overlap between the Boltzmann distributions. Thus, the methods fail when our two  target states are very dissimilar. This raises the question --- what do we do when our two states are essentially disjoint in configuration space and we would like to estimate their free energy difference? 

The classical approach to this challenge involves defining a path through phase space, controlled by a coupling parameter $\lambda$, that connects state 0 ($\lambda=0$) and state 1 ($\lambda=1$). The definition of this path allows for the sampling of a series of states $\{\lambda_0=0,\dots, \lambda_k, \dots,\lambda_{M-1}=1\}$ and the total free energy difference between states 0 and 1 can be computed as a sum over the free energy differences of adjacent states $k$ and $k+1$ determined by FEP, BAR, or MBAR, or by applying thermodynamic integration (TI) \cite{RN135}. This approach is called \emph{windowing}, and can be used to obtain low-error estimates of the free energy when the direct configuration space overlap between states 0 and 1 is essentially zero, since the configuration space overlap between states $k$ and $k+1$ can be made much higher with a rational choice of the parametric path. This approach comes with two disadvantages, the first of which is the additional computational cost of sampling the intermediate states. Another issue is that, given only samples from the path endpoints (states 0 and 1), it is not clear what the best choice of parametric path is. Efficiently determining effective paths in phase space, and their connection to non-equilibrium methods based on the Jarzynski equality \cite{RN205,RN203}, is an active area of ongoing research \cite{RN207,RN204}.

A partial resolution to these issues was proposed by Jarzynski \cite{RN151}, who showed that a differentiable, bijective mapping ($\mathcal{M}$) between the configuration spaces of states 0 and 1 can be used to greatly reduce the number of samples needed to achieve a target accuracy in a one-step free energy estimate using standard estimators (FEP or BAR). Jarzynski referred to this approach as Targeted Free Energy Perturbation (TFEP), and this method provides a way to overcome the challenge of obtaining accurate and precise free energy estimates between states that are disjoint in phase space. While very powerful in theory, this approach has provided limited utility in practice, because finding a mapping between the configuration spaces of two high-dimensional Boltzmann distributions that effectively increases phase space overlap is far from trivial. The typical approaches to finding $\mathcal{M}$ can be divided into two main categories --- (a) derivation of analytical mapping functions from physical intuition, and (b) casting the mapping search as a machine-learning problem, with the mapping being represented by a highly expressive function with trainable parameters \cite{RN182}.

The construction of analytical mappings from physical intuition is advantageous as such mappings can accelerate free energy calculations with little additional computational cost. Successful applications of TFEP using such an approach have been limited to free energy estimation problems where the useful transformations happen to be quite intuitive --- for example, estimating the free energy between crystals at different temperatures \cite{RN186,RN184}, the free energy differences between different rigid water models \cite{RN185}, or the free energy change of expanding a solute in a solvent \cite{RN151}.  A particularly informative application of this approach was provided by Hahn \& Then, where they proposed accelerating ``particle-insertion" based chemical potential calculations by applying a 1D mapping that modifies solute-solvent distances \cite{RN152}. Their heuristic approach was to use the solute-solvent mapping that led to mapped configurations from state 0 (non-interacting inserted solute) having the same solute-solvent radial distribution function as state 1 (interacting inserted solute). While intuitively such a mapping should improve overlap, their approach turned out to offer only modest reductions in standard error (we later provide a mechanistic explanation for this failing in Sec. \ref{mnfs}). This emphasizes the major problem with analytical mapping functions based on intuition --- there is no guarantee that such mappings meaningfully increase phase space overlap given the complex, many-body nature of Boltzmann distributions (especially in the liquid phase), which can lead to counterintuitive mappings being the most effective.

Casting mapping construction as a machine-learning problem alleviates the limited utility observed with intuitive, analytical mappings \cite{RN182}. A general transformation between two probability distributions of arbitrary complexity can be represented by a type of generative machine-learning model known as a \emph{normalizing flow} (NF) \cite{RN208}. These are neural networks that are differentiable and invertible by construction, and designed so that they can be trained to transform samples from one probability distribution to resemble samples from another. Theoretically, these mapping functions can be expressive enough such that they establish a perfect transformation between distributions, where mapped samples from state 0 are indistinguishable from samples from state 1. NFs that can reliably generate samples of a target Boltzmann distribution are commonly referred to as Boltzmann Generators (BGs) \cite{Noe2019boltzmann}. By training a NF to minimize the Kullback-Leibler Divergence (KLD) between a transformed and target Boltzmann distribution, NFs have been used to find complex transformations that accelerate the calculation of free energies from simulation between systems at different thermodynamic conditions ($T,P$) \cite{RN153} or with different descriptions of their interatomic interactions \cite{RN190}. While effectively accelerating the sampling of crystals \cite{RN195,RN194,RN192,RN188} and biomolecules \cite{RN187, RN193,Midgley2023flow,RN196}, NFs have struggled to realize substantial gains for even simple liquid systems \cite{Schebek2025solvation,RN183}. NFs are difficult and costly to train, and are typically only applicable to the specific systems the model was trained on, making such approaches outperform direct MD/MC sampling only in some select cases. Extensions have attempted to increase the transferability and scaling of these methods with system size, but these typically involve either more intensive training procedures or inference-time corrections, with arbitrarily large and diverse systems still being out of reach for any given trained model \cite{RN193,RN153,schebek2025scalable}.

Considering the limited effectiveness of the intuition-based mappings, and the transferability and efficiency issues of NFs/BGs, we propose here a novel combination of the two approaches called a ``Minimal" Normalizing Flow (MNF). This involves constructing a NF that is low-dimensional but transforms the system in a way that is likely to increase phase space overlap while being purposefully non-expressive (incapable of achieving a perfect mapping). This limits the number of training parameters by orders of magnitude, and allows for the construction of NF architectures that can be trained in seconds to minutes on a single GPU, and are transferable across system sizes. Previous work has considered alternative losses to the KLD and raised concerns about its effectiveness, especially for unidirectional training of NFs \cite{Midgley2023flow,Felardos2023designing}. Here, we rigorously show why the KLD fails as an optimization target for non-expressive flows between two Boltzmann distributions and propose a novel training scheme that resolves this issue. Together, these theoretical advances provide a simple methodology to obtain useful targeted free energy estimates at negligible additional computational cost. We demonstrate the utility of our method by applying it to accelerate the calculation of free energy differences in several complex solvation scenarios, specifically the thermodynamics of a pure Lennard-Jones (LJ) fluid, the phase behavior of a binary LJ mixture, and the hydration of sodium and chloride ions in water.

The remainder of this paper is organized as follows. In Sec. \ref{background}, the relevant theoretical background and derivations relevant to training MNFs are presented. In Sec. \ref{methodology}, we elaborate on the details of the model systems we examine including how MD simulations are run and how MNFs are trained. In Sec. \ref{results}, free energy calculations for the pure LJ fluid, binary LJ mixture, and ion hydration test systems are presented. We show how our novel training strategy outperforms the identity mapping, training using the KLD, and the heuristic-based mappings of Hahn \& Then in accelerating chemical potential calculations for LJ systems, all at negligible computational cost. We also present, what is to the best of our knowledge, the first application of targeted free energy estimation to electrolyte systems. In Sec. \ref{discussion}, we analyze the performance of our method, discussing possible weaknesses and future extensions, and conclusions from these results are drawn. 

\section{Theory}
\label{background} 

\subsection{Free Energy Estimation}
Let us begin by assuming that we are attempting to estimate the free energy difference between two states at \emph{NVT} conditions. We shall refer to these two ensembles as state 0 and state 1, assuming that they only differ in the definition of their potential energy functions ($U_0$ and $U_1$). The configuration space densities of these two states are given by $\rho_0(x)=e^{-\beta U_0(x)}/Z_0$ and $\rho_1(y)=e^{-\beta U_1(y)}/Z_1$, where we have that --- 
\begin{equation}
    Z_i=\int\cdots\int e^{-\beta U_i\left(\textbf{r}_1, \dots,\textbf{r}_N\right)}d\textbf{r}_1 \dots d\textbf{r}_N\
\end{equation}
is the partition function of ensemble $i$, using the shorthand notation of $x$ or $y$ referring to a specific configuration of particles $\{\textbf{r}_1, \dots,\textbf{r}_N\}$, and $\beta=1/(k_BT)$ is the inverse temperature \cite{McQuarrie2000}. The free energy difference between these states is given by $\Delta F^{0\rightarrow1}=-k_B T\ln\left(Z_1/Z_0\right)$.
Our goal is to obtain an estimate of the free energy difference ($\Delta \hat{F}^{0\rightarrow1}$) given that we collected independent samples of one or both of our states from a MC or MD simulation. 

This estimate is entirely determined by the work of infinitely fast switching between ensembles in either direction, given by ---
$W^{(F)}(x) = U_1(x)-U_0(x)$ and 
$W^{(R)}(y) = U_0(y)-U_1(y)$,
where $x\in\{x_i\}_{i=1}^{N_{\text{ind}}}$ are samples from state 0 and $y\in\{y_i\}_{i=1}^{N_{\text{ind}}}$ are samples from state 1, and $N_{\text{ind}}$ is the number of independent samples. These work terms can be interpreted as the infinitely fast limit of the work done by a non-equilibrium process that starts in one ensemble and transforms it into the other. Thus, each state generates a probability distribution of possible values of this work ($\mathbb{P}_0\left(W^{(F)}\right)$ for state 0 and $\mathbb{P}_1\left(-W^{(R)}\right)$ for state 1, where the argument is negated so that we are considering the ``work done on the system" in both cases). The equilibrium version of the Crooks Fluctuation Theorem (CFT) \cite{RN206} provides us with an important relationship between these probabilities ---
\begin{equation}
    \label{eq:cft}
\frac{\mathbb{P}_0\left(W\right)}{\mathbb{P}_1\left(-W\right)}=e^{\beta\left(W-\Delta F^{0\rightarrow1}\right)}
\end{equation}
The CFT tells us that forward and reverse work distributions have to intersect at $W=\Delta F^{0\rightarrow1}$. This expression allows us to construct free energy estimates from sets of simulated samples of work values. For example, rearranging Eq. \ref{eq:cft} and integrating both sides gives us the Free Energy Perturbation (FEP) or Zwanzig estimator \cite{RN110} --- 
\begin{equation}
\begin{aligned}
\label{eq:fep}
\Delta\hat{F}^{0\rightarrow1}=-k_BT\ln\left<e^{-\beta W^{(F)}(x)}\right>_0 =k_BT\ln\left<e^{-\beta W^{(R)}(y)}\right>_1
\end{aligned}
\end{equation}
which provides a free energy estimate when we have sampled only one of our states. However, this estimate is prone to high bias and variance if the distributions $\mathbb{P}_0\left(W\right)$ and $\mathbb{P}_1\left(-W\right)$ don't sufficiently overlap. 

Our free energy estimate can be significantly improved if we instead collect samples from both state 0 and state 1. Shirts \emph{et al.} showed that this allows Eq. \ref{eq:cft} to be interpreted as a logistic regression problem with a free parameter $\Delta F^{0\rightarrow1}$ \cite{RN210}. In this framing, the maximum-likelihood estimate of $\Delta F^{0\rightarrow1}$ is given by the Bennett Acceptance Ratio (BAR) \cite{RN18} ---
\begin{equation}
\begin{aligned}
\label{eq:bar}
    \left<\frac{1}{1+\exp\left(
    \beta\left[W^{(F)}(x)-\Delta\hat{F}^{0\rightarrow1}\right]\right)}\right>_0 = \left<\frac{1}{1+\exp\left(
    \beta\left[W^{(R)}(y)+\Delta\hat{F}^{0\rightarrow1}\right]\right)}\right>_1
\end{aligned}
\end{equation}
where we have assumed that the same number of samples was collected from both states. The equation above  is self-consistently solved for $\Delta\hat{F}^{0\rightarrow1}$. This is the provably lowest variance estimate of $\Delta F^{0\rightarrow1}$, with the asymptotic per-sample variance (from the Cramer-Rao lower bound) given by --- 
\begin{equation}
\label{eq:bar_var}\sigma^2\left(\beta\Delta\hat{F}^{0\rightarrow1}\right)=\frac{2}{N_{\text{ind}}}\left[\frac{1}{\mathcal{H}_{0,1}}-1\right]
\end{equation}
where we assume that $N_{\text{ind}}$ samples are collected from each state and $\mathcal{H}_{0,1}$ is the harmonic overlap metric between states 0 and 1 --- 
\begin{equation}
\begin{aligned}
\mathcal{H}_{0,1}&=\int\dots\int\frac{2\rho_0(\textbf{r}^N)\rho_1(\textbf{r}^N)}{\rho_0(\textbf{r}^N)+\rho_1(\textbf{r}^N)}d\textbf{r}_1\dots d\textbf{r}_N\\ &= \int\frac{2\mathbb{P}_0\left(W\right)\mathbb{P}_1\left(-W\right)}{\mathbb{P}_0\left(W\right)+\mathbb{P}_1\left(-W\right)}dW
\end{aligned}
\end{equation}
We can see that this expression measures the degree of configuration space overlap between our Boltzmann distributions, and for a given number of samples $N_{\text{ind}}$ the (asymptotic) variance in our free energy estimate is inversely proportional to $\mathcal{H}_{0,1}$. This identifies a general problem --- if $\mathcal{H}_{0,1}\approx0$ (states 0 and 1 barely overlap in phase space), then the number of samples $N_{\text{ind}}^*=\left(2/\sigma^2_\text{goal}\right)\left[\mathcal{H}_{0,1}^{-1}-1\right]$ required to achieve a target accuracy $\sigma^2_\text{goal}$ may be computationally infeasible to generate from simulation. A typical approach to this problem is windowing --- inserting intermediate states between 0 and 1 that are likely to increase mutual configuration space overlap, but the optimal choice of windows cannot be known \emph{a priori}. An alternative resolution to this challenge is \emph{targeted} free energy estimation. 

\subsection{Targeted Free Energy Estimation (TFEE)}

Jarzynski showed that the configuration space overlap between the Boltzmann distributions of states 0 and 1 can be enhanced by a bijective, differentiable mapping between the configuration spaces of our two systems \cite{RN151}. This mapping, $\mathcal{M}:x\rightarrow\phi(x)=y$ attempts to change the coordinates of configurations $x$ sampled from $\rho_0(x)$ such that they resemble likely coordinates $y$ sampled from $\rho_1(y)$. Since bijectivity ensures invertability, the inverse mapping can be applied to configurations $y$ as well. The mapping ``warps" the phase space densities, giving the following mapped versions --- $\tilde{\rho}_1(y)=\rho_0(x)/|\text{det }\mathcal{J}(x)|$ (0 mapped to 1) and $\tilde{\rho}_0(x)=\rho_1(y)/|\text{det }\mathcal{J}^{-1}(y)|$ (1 mapped to 0). Thus, the mapping also ``warps" the forward and reverse work values, giving the following definition of ``mapped works" ---
\begin{equation}
\begin{aligned}
    \label{eq:wf_mapped}
    \tilde{W}^{(F)}(x) =  U_1\left(\phi(x)\right) - U_0(x) -k_B T\ln\left|\text{det }\mathcal{J}(x)\right|
\end{aligned}
\end{equation}
\begin{equation}
\begin{aligned}
    \label{eq:wr_mapped}
    \tilde{W}^{(R)}(y) = U_0\left(\phi^{-1}(y)\right) - U_1(y) -k_BT\ln\left|\text{det }\mathcal{J}^{-1}(y)\right|
\end{aligned}
\end{equation}
where $\mathcal{J}=\partial\phi(\textbf{r}_1,\dots,\textbf{r}_N)/\partial(\textbf{r}_1,\dots,\textbf{r}_N)$ is the Jacobian of our transformation. 

Hahn and Then showed that the probabilities of observing these ``mapped works" follow a generalized version of the CFT in Eq. \ref{eq:cft} --- $\mathbb{P}_0(\tilde{W})=\mathbb{P}_1(-\tilde{W})e^{\beta\left(\tilde{W}-\Delta F^{0\rightarrow1}\right)}$, where the mapped work distributions can also be written as $\mathbb{P}_0(\tilde{W})=\mathbb{P}_0(W|\mathcal{M})$ \cite{RN152}. This generates ``targeted" analogues of our FEP (Eq. \ref{eq:fep}) and BAR (Eq. \ref{eq:bar}) estimators, where the standard forward and reverse works are replaced with Eqs. \ref{eq:wf_mapped} and \ref{eq:wr_mapped}, respectively. The ``mapped" overlap between our Boltzmann distributions can then be computed as --- 
\begin{equation}
\begin{aligned}
\tilde{\mathcal{H}}_{0,1}= \int\frac{2\mathbb{P}_0(W|\mathcal{M})\mathbb{P}_1(-W|\mathcal{M})}{\mathbb{P}_0(W|\mathcal{M})+\mathbb{P}_1(-W|\mathcal{M})}dW
\end{aligned}
\end{equation}
where a useful mapping is one such that $\tilde{\mathcal{H}}_{0,1}> \mathcal{H}_{0,1}$. This allows us to define the following metric of improvement that a targeted free enery estimate provides --- 
\begin{equation}
\tilde{n}=\left(\frac{\sigma_{\text{id}}}{\sigma_{\text{targ}}}\right)^2
\end{equation}
This definition comes from applying a re-arranged form of the asymptotic per-sample variance bound in Eq. \ref{eq:bar_var}, where $\sigma_{\text{id}}$ is the standard error in the unmapped free energy estimate, and $\sigma_{\text{targ}}$ is the standard error in the targeted free energy estimate. The quantity $\tilde{n}$ gives a measure of how many times fewer samples need to be collected for the targeted estimate to achieve the same error as the standard one.

A surprising result is that if we can find a mapping that perfectly maps one ensemble onto the other, our mapped work distributions become Dirac delta functions centered at $\Delta F^{0\rightarrow1}$, and thus our free energy estimate can be determined exactly from a single sample. However, this is all complicated by the fact that finding useful mappings is very difficult. 

\subsection{Learned Mappings}
\label{sec:learned_map}
Determining a useful mapping $\phi(x)$ is often cast as a machine learning problem, where the mapping is represented by a maximally expressive NF \cite{RN182}. Thus, our mapping becomes $\phi(x, \boldsymbol{\theta})$ where $\boldsymbol{\theta}$ is some vector of trainable parameters. To train our mapping, we need a metric of how close the ``mapped" version of an ensemble is to the true distribution it is attempting to match. The general form of such a metric is called the $f$-divergence, which measures the differences between probability distributions. An $f$-divergence is defined as --- 
\begin{equation}
\label{eq:f_div_0}
    \mathbb{D}_f(g||h)=\int g(x)f\left(\frac{h(x)}{g(x)}\right)dx
\end{equation}
where $f$ is a convex function satisfying $f(1)=0$. The state-of-the-art choice of $f$-divergence is the Kullback-Leibler Divergence (KLD). For our purposes, the KLD is taken between the mapped ensemble 0 density, and the true ensemble 0 density (forward direction, $f(t)=-\ln t$) ---
\begin{equation}
\begin{aligned}
\label{eq:kld_fwd}
\mathbb{D}^{(F)}_{\text{KL}}\left(\rho_0||\tilde{\rho}_0\right) &= \int \rho_0(x)\ln\left(\frac{\rho_0(x)}{\tilde{\rho}_0(x)}\right)dx \\&= \beta\left<\tilde{W}^{(F)}(x)\right>_0 - \beta\Delta F^{0\rightarrow1}
\end{aligned}
\end{equation}
It is worth noting that we call this the ``forward" direction due to the involvement of the thermodynamic work of the forward process, while the machine learning community often refers to this as the ``reverse" direction, based on the direction the mapping is acting. We proceed similarly for ensemble 1 (what we call the reverse direction) we obtain --- 
\begin{equation}
\label{eq:kld_rev}
\mathbb{D}^{(R)}_{\text{KL}}\left(\rho_1||\tilde{\rho}_1\right) = \beta\Delta F^{0\rightarrow1} + \beta\left<\tilde{W}^{(R)}(y)\right>_1
\end{equation}
If we notice that $\Delta F^{0\rightarrow1}$ is independent of $\boldsymbol{\theta}$, then Eqs. \ref{eq:kld_fwd} and \ref{eq:kld_rev} can be minimized with respect to $\boldsymbol{\theta}$ without knowledge of $\Delta F^{0\rightarrow1}$ in unidirectional training. The KLD is also commonly employed in bidirectional training, where the KLDs in each direction are added and the value of $\Delta F$ cancels producing the loss function ---
\begin{equation}
\begin{aligned}
\label{eq:loss_kl}
 \mathcal{L}_{\text{KL}}(\boldsymbol{\theta}) &= \mathbb{D}^{(F)}_{\text{KL}}\left(\rho_0||\tilde{\rho}_0\right)+\mathbb{D}^{(R)}_{\text{KL}}\left(\rho_1||\tilde{\rho}_1\right) \\&= \beta\left<\tilde{W}^{(F)}(x,\boldsymbol{\theta})\right>_0+\beta\left<\tilde{W}^{(R)}(y,\boldsymbol{\theta})\right>_1
\end{aligned}
\end{equation}
This loss is typically minimized using stochastic optimization techniques where the ensemble averages in Eq. \ref{eq:loss_kl} are taken directly over gathered simulation samples at each training step. The minimization of this loss produces the optimal set of mapping parameters $\boldsymbol{\theta}^*$, which generate a useful mapping. 

Applying Jensen's inequality to the mapped forward and reverse Zwanzig estimators, we get the following inequality ---
\begin{equation}
\label{eq:jensen}
-\left<\tilde{W}^{(R)}(y,\boldsymbol{\theta})\right>_1\leq\Delta F^{0\rightarrow1} \leq \left<\tilde{W}^{(F)}(x,\boldsymbol{\theta})\right>_0
\end{equation}
This means that minimizing Eq. \ref{eq:loss_kl} amounts to finding the set of mapping parameters that minimizes the differences in the means of the mapped forward ($\mathbb{P}_0(W|\mathcal{M})$) and mapped reverse ($\mathbb{P}_1(-W|\mathcal{M})$) work distributions, where the entire distributions are modified as to push the reverse distribution mean towards $\Delta F^{0\rightarrow1}$ from below and the forward distribution mean towards $\Delta F^{0\rightarrow1}$ from above. 

This describes the standard approach to training normalizing flows in chemical physics, where $\phi(x,\boldsymbol{\theta})$ is designed to be maximally expressive, i.e. capable of perfectly mapping from one ensemble to the other ($\rho_i=\tilde{\rho}_i$), and the KLD training aims at achieving this perfect mapping by satisfying the equality --- $\left<-\tilde{W}^{(R)}(y,\boldsymbol{\theta})\right>_1=\Delta F^{0\rightarrow1} = \left<\tilde{W}^{(F)}(x,\boldsymbol{\theta})\right>_0$. Due to the complexity of the optimization problem, this is rarely achieved in practice, but useful mappings are still typically found. 
\subsection{``Minimal" Normalizing Flows}
\label{mnfs}
The present work diverges from much of the previous literature by still casting mapping determination as a machine-learning problem, while relaxing the constraint that $\phi(x,\boldsymbol{\theta})$ needs to be maximally expressive. Instead, we propose the following --- flows can be built that are intentionally non-expressive, but operate on low-dimensional, physically informed coordinates that are most likely to increase phase space overlap between states 0 and 1. We refer to such a flow as a \emph{``Minimal" Normalizing Flow} (MNF). The central utility of MNFs is that their simplicity allows for training ``on-the-fly" at negligible computational cost, while still enhancing phase space overlap between our systems of interest. 

In this work we particularly consider TFEE applied to particle-insertion free energy calculations. In such calculations, the interactions of a non-interacting species in a fluid are turned on, and the free energy of this process is computed. This free energy is a finite difference approximation of the chemical potential of the ``inserted" species --- 
\begin{equation}
\begin{aligned}
    \mu_{i} = \left(\frac{\partial G}{\partial N_{i}}\right)_{N_{j\neq i},P,T} &\approx \frac{G_{N_{i}} - G_{N_{i}-1}} {N_{i} - (N_{i}-1)}  \\ &= \Delta G^{N_{i}-1\rightarrow N_{i}}
\end{aligned}
\end{equation}
where $N_i$ is the number of species $i$ in the system, $P$ is the pressure, $G$ is the Gibbs free energy, and $\mu_i$ is the chemical potential of species $i$.

\begin{figure*}[t!]
\centering
\includegraphics[width=16cm]{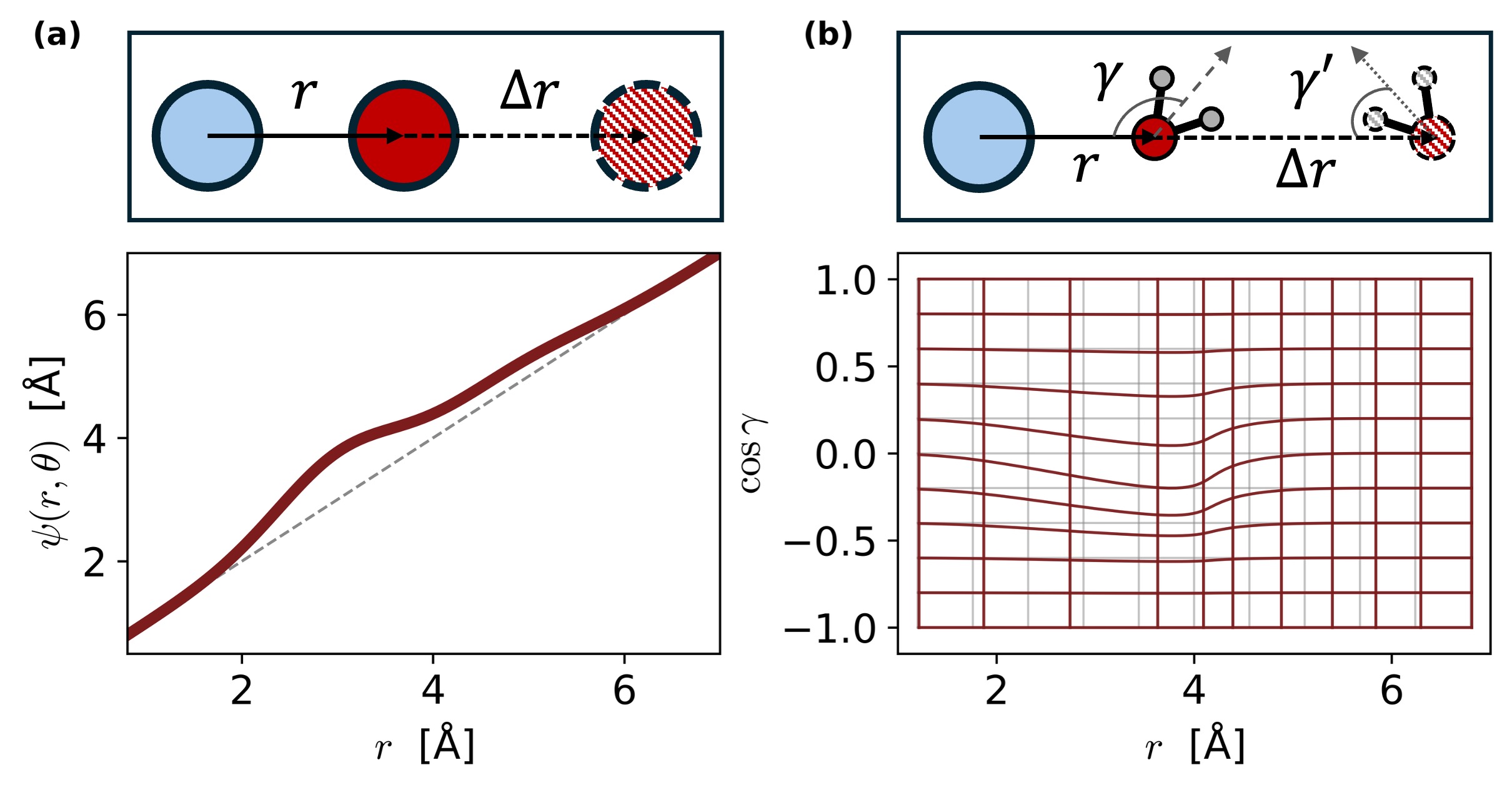}
\caption{\textbf{Demonstration of 1D and 2D Minimal Normalizing Flows}. (a) A one-dimensional MNF, applied uniformly to every solvent particle a distance $r$ from the inserted solute. The solvent particle is mapped to a new radial distance $r'=\psi(r)=r+\Delta r$. (b) A two-dimensional MNF, where every oxygen at distance $r$ away from the inserted ion is mapped to a distance $r'=\psi(r)=r+\Delta r$ while the water molecule is rotated as to modify the angle between its H-O-H bisector and the ion-O vector to be $\gamma'=g(\gamma|r)$. The grey grid in the bottom panel represents a uniform spacing in $(r,\cos\gamma)$ space, and the red grid shows how our 2D MNF would deform this 2D plane obtaining new points $(r',\cos\gamma')$.}
\label{fig:flow_demo}
\end{figure*}

In such ``particle insertion" scenarios, the natural, low-dimensional coordinate differentiating the ``non-inserted" and ``inserted" states is the radial distance of all solvent atoms from the solute. Thus, we can propose a MNF that is a 1-dimensional radial map, defined as $\phi(\textbf{r}_{\text{ins}},\textbf{r}_1,\dots,\textbf{r}_{N-1})=\{\textbf{r}_{\text{ins}},f(\textbf{r}_1,\textbf{r}_{\text{ins}}), \dots, f(\textbf{r}_{N-1},\textbf{r}_{\text{ins}})\}$, where -
\begin{equation}
\label{eq:radial_map}
    f (\textbf{r}_i,\textbf{r}_{\text{ins}})=\frac{\textbf{r}_i-\textbf{r}_{\text{ins}}}{\norm{\textbf{r}_i-\textbf{r}_{\text{ins}}}}\psi\left(\norm{\textbf{r}_i-\textbf{r}_{\text{ins}}}\right) 
\end{equation}
where $\textbf{r}_{\text{ins}}$ is the position of the inserted particle, and $\psi:\mathbb{R}\rightarrow\mathbb{R}$ is a bijective, differentiable, radial mapping function, that modifies the distance of each solvent atom from the solute. An example of such a function and a schematic of its behavior are shown in the left panel of Fig. \ref{fig:flow_demo}. The log of the Jacobian determinant of this transformation takes the following simple analytical form --- 
\begin{equation}
    \ln \left|\text{det } \mathcal{J}(x)\right| = \sum_{i\neq\text{ins}}\ln\left[\frac{\psi(r_{i,\text{ins}})^2\psi'(r_{i,\text{ins}})}{r_{i,\text{ins}}^2}\right]
    \label{eq:jac_1d}
\end{equation}
where $r_{i,\text{ins}} = \norm{\textbf{r}_i-\textbf{r}_{\text{ins}}}$. 

Such a mapping was first studied by Hahn and Then, where they defined $\psi$ such that the mapped solute-solvent radial distribution function (RDF) of state 0 matched the state 1 solute-solvent RDF \cite{RN152}. They showed that this choice was somewhat effective for accelerating chemical potential calculations of a pure LJ fluid modeling Argon. While this is an intuitive and efficient scheme, we can see why this is far from optimal by examining the theoretical form of the KLD for a radial mapping (Eq. \ref{eq:radial_map}) applied to a pure fluid with a single ``inserted" particle ---  
\begin{equation}
\begin{aligned}
\label{eq:cf_decomp}
&\mathbb{D}^{(F)}_{\text{KL}}\left(\rho_0||\tilde{\rho}_0\right) \propto \left<\beta U_1(\phi(x))-\ln\left|\text{det }\mathcal{J}(x)\right|\right>_0 \\&= 4\pi\rho\int\left[\beta u_1(\psi(r)) -\ln\left[\frac{\psi(r)^2\psi'(r)}{r^2}\right]\right]g^{(2)}_0(r)r^2dr  \\ &+ 4\pi^2\rho^2\beta \int\int\int u_1\left(\sqrt{\psi(r)^2+\psi(r')^2-2\psi(r)\psi(r')\cos\theta}\right)g^{(3)}_0(r, r', \theta) r^2r'^2\sin(\theta)d\theta dr dr'
\end{aligned}
\end{equation}
where our ``inserted" particle is centered at the origin, $\rho=N/V$ is the fluid number density, $u_i(r)$ is the pair potential of state $i$, and $g_i^{(n)}$ is the $n$-body correlation function in ensemble $i$ between the ``inserted" particle, and $n-1$ solvent particles. A derivation of the above expression can be found in Sec. 2 of the supplementary information (SI), with the reverse KLD treated analogously. In the above expression we have discarded the $-\left<U_0(x)\right>_0-\Delta F$ terms from $\mathbb{D}^{(F)}_{\text{KL}}\left(\rho_0||\tilde{\rho}_0\right)$ since they are independent of $\psi$. It is clear from the second term of Eq. \ref{eq:cf_decomp} that the RDF-matching definition of $\psi$ does not optimally maximize the phase space overlap between our states --- this heuristic does not consider the induced three-body effect (on solvent-solvent distances) caused by the radial mapping. Even though simple heuristics such as RDF-matching carry intuitive force for low-dimensional, physically informed mappings (MNFs), we argue that the minimization of a statistical distance metric between mapped and true ensembles is a more reliable way to obtain useful MNFs.

The re-written KLD in Eq. \ref{eq:cf_decomp} prompts us to ask two questions --- does this form provide us with an alternative optimization approach? If so, is the KLD the right loss function for a MNF?

\subsubsection{Correlation Function Decompositions}
\label{sec:cfd}

As mentioned before, typical determination of the optimal set of mapping parameters is done through stochastic optimization. However, Eq. \ref{eq:cf_decomp} offers us a route to deterministic training. If $g^{(2)}$ and $g^{(3)}$ are pre-computed for states 0 and 1 prior to training, Eq. \ref{eq:cf_decomp} can be evaluated as a deterministic integral expression at each training step. Due to the elimination of ensemble averaging and energy evaluations over entire configurations, this has  the potential to be more efficient than stochastic optimization.

We can see that a general $f$-divergence (as defined by Eq. \ref{eq:f_div_0}) between true and mapped ensembles takes the form (taking the forward direction as an example) -
\begin{equation}
\begin{aligned}
\label{eq:f_div}
    \mathbb{D}_f(\rho_0 || \tilde{\rho}_0)&=\int \rho_0(x)f\left(\frac{\tilde{\rho}_0(x)}{\rho_0(x)}\right)dx \\ 
    &=\left<f\left(e^{\beta\left[\Delta F- W^{(F)}(x)\right]}\right)\right>_0 =\left<f\left(e^{\mathcal{W}}\right)\right>_0
\end{aligned}
\end{equation}
where for shorthand we have defined $\mathcal{W}=\beta\left[\Delta F- W(x)\right]$. We now note two general facts --- 
\begin{itemize}
    \item \underline{Proposition 1}: If $F(\mathcal{W}) = f\left(e^{\mathcal{W}}\right)$ is a polynomial function of $\mathcal{W}$ of degree $k$, and we are applying a 1D radial map (as defined in Eq. \ref{eq:radial_map}), then we need to pre-compute correlation functions up to $(2k+1)$-body correlations in order to resolve an integral representation of $\mathbb{D}_f$ analogous to Eq. \ref{eq:cf_decomp}. If $F(\mathcal{W}) =f\left(e^{\mathcal{W}}\right)$ is a non-polynomial function of $\mathcal{W}$, all $N$-body correlations are required.
    \item \underline{Proposition 2}: The KLD is the only $f$-divergence that for Boltzmann distributions results in an ensemble average of a function that is linear ($k=1$) in $\mathcal{W}$.
\end{itemize}
Proposition 1 can be proven by examining the derivation of Eq. \ref{eq:cf_decomp} in Sec. 2 of the SI. Evaluating $\left<\mathcal{W}^k\right>_0$ will involve averages of the product of $k$  different solvent-solvent energy pairs, influenced by the 1D map, which involves $2k$ distinct particles plus the inserted solute, requiring $g^{(2k+1)}$ to evaluate the ensemble average. Non-polynomial expressions can be viewed in polynomial form as a Taylor series, and thus require all $N$-body correlations. Proposition 2 is immediately clear from Eq. \ref{eq:f_div}.  

The above propositions make it clear that while the correlation function-based approach is theoretically rich, it is only feasible as a training strategy when optimizing the KLD for a 1D radial mapping. In this context, we only need to precompute correlation functions up to three-body correlations. In any more complex scenario, higher order correlations are needed, and these are very expensive to obtain from simulation. This begs the question --- is the KLD an effective loss for training intentionally non-expressive mappings? 

\subsubsection{Is the KLD an effective loss function for MNFs?}
\label{sec:kld_pathology}

We argue that the KLD is the incorrect optimization target for MNF training. The reason for this is as follows --- in Sec. \ref{sec:learned_map}, we showed that the minimization of the KLD amounts to minimizing the distance between the means of the forward and reverse mapped work distributions. For a fully expressive mapping, this objective typically coincides with increasing the phase space overlap between our two systems. This is because the perfect mapping results in $\tilde{\mathcal{H}}_{0,1}=1$ and $\mathbb{P}_0(W|\mathcal{M}) = \mathbb{P}_1(-W|\mathcal{M})=\delta(W-\Delta F)$ as per the mapped CFT. However, we propose MNFs as intentionally non-expressive maps, that means that these mappings are incapable of pushing the means of the mapped work distributions arbitrarily close together. 

If we determine our MNF parameters as $\boldsymbol{\theta}^*=
\text{argmin}_{\boldsymbol{\theta}} \mathcal{L}_{\text{KL}}(\boldsymbol{\theta})$, then our MNF will push the work distribution means as close as possible to each other. However, there is no constraint on how this mapping affects $\tilde{\mathcal{H}}_{0,1}$ and thus how significantly overlap is enhanced, since the KLD depends only on the 1st moment of our mapped work distributions. Given a highly non-expressive map, this means that we can achieve a KLD minimum \emph{without enhancing phase space overlap at all}. Therefore, we argue that the KLD is the wrong optimization target for MNF training, as it does not target phase space overlap directly. 

Combined with the claims of Sec. \ref{sec:cfd} (propositions 1 and 2), this allows us to construct a cohesive argument for how MNF training should be approached --- 
\begin{itemize}
    \item Any loss that explicitly targets overlap is not the KLD.
    \item Any loss that explicitly targets overlap is thus nonlinear in $\mathcal{W}$.
    \item In order to employ correlation function-based optimization on an analogue of Eq. \ref{eq:cf_decomp} for this overlap-targeting loss, we would require the calculation of at least five-body correlation functions.
    \item Therefore, correlation function-based optimization is intractable for any overlap metric, and we must employ direct stochastic optimization to train useful MNFs.
\end{itemize}
The remaining design question is --- which overlap targeting loss function should we optimize?

\subsection{A New Loss Function}

As we have noted before, the asymptotic variance in our free energy calculations is proportional to $\tilde{\mathcal{H}}_{0,1}^{-1}$, so we seek the parameters $\boldsymbol{\theta}^*$ that maximize $\tilde{\mathcal{H}}_{0,1}$ and thus minimize our variance. It is worth noting that $1-\tilde{\mathcal{H}}_{0,1}$ is an $f$-divergence, and minimizing this is an equivalent optimization problem to minimizing what we will refer to as the ``harmonic distance" (HD) --- 
\begin{equation}
\begin{aligned}
    \mathbb{D}_{\text{HD}}&= -\ln\tilde{\mathcal{H}}_{0,1} \\ &= -\ln\left<\frac{2}{1+e^{-\mathcal{W}}}\right>_0 = -\ln\left<\frac{2}{1+e^{\mathcal{W}}}\right>_1
\end{aligned}
\end{equation}
The advantage of using this as a loss function to train our mapping is that we are directly maximizing the phase space overlap relevant to the quality of our free energy estimate, as per Eq. \ref{eq:bar_var}. However, the major disadvantage of using the HD as a loss is that its behavior depends on $\Delta F^{0\rightarrow1}$, the very quantity we are attempting to estimate. If our initial estimate of $\Delta F^{0\rightarrow1}$ is very biased (which is especially relevant when we are doing one-sided estimation with FEP), then the optimality condition of our optimization is seriously polluted and we will not find the optimal map. This point is elaborated on in Sec. 3.1 of the SI.

To alleviate this problem, we propose to use a new loss function --- the Bhattacharyya Distance (BD) between our mapped work distributions ---  
\begin{equation}
\begin{aligned}
    \mathbb{D}_{\text{BD}}&= - \ln \mathbb{B}_{0,1} \\
    &= -\ln\left(\int\sqrt{\mathbb{P}_0(\tilde{W})\mathbb{P}_1(-\tilde{W})}d\tilde{W}\right) \\ &= -\ln\left<e^{\frac{\mathcal{W}}{2}}\right>_0 = -\ln\left<e^{-\frac{\mathcal{W}}{2}}\right>_1 
\end{aligned}
\end{equation}
where $\mathbb{B}_{0,1}$ is referred to as the Bhattacharyya Coefficient (BC) \cite{MR10358}. The BD is simply a different overlap metric (based on the geometric distribution mean) that approaches 0 in the limit of perfect overlap, just as the HD. We can see that the BC serves as an upper bound to $\tilde{\mathcal{H}}_{0,1}$, where we always have that $\tilde{\mathcal{H}}_{0,1}\leq\mathbb{B}_{0,1}$ and thus $\mathbb{D}_{\text{BD}}\leq\mathbb{D}_{\text{HD}}$. While this loss has the disadvantage of not controlling $\tilde{\mathcal{H}}_{0,1}$ directly, we can see that its behavior is actually independent of our estimate of the free energy. Considering the forward direction as an example, we have --- 
\begin{equation}
\begin{aligned}
    \mathbb{D}_{\text{BD}} &= -\ln\left<e^{\frac{\beta}{2}\left(\Delta F - W^{(F)}(x)\right)}\right>_0 \\ &= -\frac{1}{2}\beta\Delta F -\ln\left<e^{-\frac{\beta}{2}W^{(F)}(x)}\right>_0 \\ 
        &\propto -\ln\left<e^{-\frac{\beta}{2}W^{(F)}(x)}\right>_0
\end{aligned}
\end{equation}
thus, the free energy is simply an additive factor, and the optimal mapping parameters that minimize $\mathbb{D}_{\text{BD}}$ do not depend on it. 

Another interesting property is that the gradients of $\mathbb{D}_{\text{HD}}$ become very noisy in the limit of negligible overlap (where the work distributions essentially have disjoint support), while this occurs to a lesser degree for $\mathbb{D}_{\text{BD}}$. In practice, both losses tend to have non-negligible gradients that tend towards the proper minimum in our numerical experiments, but this theoretical claim supports $\mathbb{D}_{\text{BD}}$ having generally more robust gradient behavior. A theoretical demonstration of this claim is presented in Sec. 3.2 of the SI. 

Lastly, it is worth noting that the functional form of the BD resembles the Jarzynski equality expression --- $\beta\Delta F=-\ln\left<e^{-\beta W^{(F)}(x)}\right>_0$ \cite{RN205}. In the BD case, the loss is non-constant due to the factor of $1/2$ in the exponential. The typical Jarzynski exponential average is very noisy, however the factor of 1/2 in the BD effectively \emph{increases the temperature} in the Boltzmann factor, decreasing the variance in the ensemble average. Based on this ``effective temperature" rationale, decreasing the value of the fraction in the exponential (which we refer to as $\alpha$) for this type of loss can help control the variance in estimating the exponential average from one of the states. This points to the existence of a generalized Bhattacharyya distance, which we refer to as a skewed Bhattacharyya distance (SBD), defined as --- 
\begin{equation}
\begin{aligned}
\mathbb{D}_{\text{BD}}^{(\alpha)} &= -\ln\left(\int\left[\mathbb{P}_0(\tilde{W})\right]^{1-\alpha}\left[\mathbb{P}_1(-\tilde{W})\right]^{\alpha}d\tilde{W}\right)\\
& = -\ln\left<e^{\alpha \mathcal{W}}\right>_0 = -\ln\left<e^{-(1-\alpha)\mathcal{W}}\right>_1
\end{aligned}
\end{equation} 
However, the lower bound $\mathbb{D}_{\text{BD}}^{(\alpha)}\leq \mathbb{D}_{\text{HD}}$ is only valid for all possible work distributions if $\alpha=1/2$, which motivates the choice of the standard Bhattacharyya distance $\left(\mathbb{D}_{\text{BD}}^{(1/2)}\right)$ as our new loss function. We expand on this fact in the SI. For this reason, along with the variance damping effects of $\alpha=1/2$, we favor the BD to alternative losses considered in the literature related to higher-$\alpha$ members of the SBD family \cite{Midgley2023flow,Felardos2023designing}.

We can construct practical losses from $\mathbb{D}_{\text{BD}}$ and $\mathbb{D}_{\text{HD}}$ by adding the ensemble average expressions from each state --- 
\begin{equation}
 \label{eq:bd_loss}
\mathcal{L}_{\text{BD}}\left(\boldsymbol{\theta}\right)= -\ln\left<e^{-\frac{\beta}{2}W^{(F)}\left(x,\boldsymbol{\theta}\right)}\right>_0-\ln\left<e^{\frac{\beta}{2}W^{(R)}\left(y,\boldsymbol{\theta}\right)}\right>_1
\end{equation}
\begin{equation}
\begin{aligned}
\mathcal{L}_{\text{HD}}\left(\boldsymbol{\theta},\Delta \hat{F}^{0\rightarrow1}\right)= -\ln\left<\frac{2}{1+e^{-\mathcal{W}\left(x,\boldsymbol{\theta},\Delta \hat{F}^{0\rightarrow1}\right)}}\right>_0 -\ln\left<\frac{2}{1+e^{\mathcal{W}\left(y,\boldsymbol{\theta},\Delta \hat{F}^{0\rightarrow1}\right)}}\right>_1
\end{aligned}
 \label{eq:hd_loss}
\end{equation}
even though the two terms in each expression are formally equal, they allow us to use all collected samples from states 0 and 1 for training. Thus, when we refer to using the KLD, BD, or HD losses, we refer to the implementation of Eqs. \ref{eq:loss_kl}, \ref{eq:bd_loss}, and \ref{eq:hd_loss} respectively. The theoretical considerations above motivate the following simple optimization strategy for any general flow where expressivity is not guaranteed, and samples are collected from both states --- 
\begin{enumerate}
    \item Train for $n_{\text{epochs, BD}}$ using $\mathcal{L}_{\text{BD}}(\boldsymbol{\theta})$ (Eq. \ref{eq:bd_loss}) to obtain $\boldsymbol{\theta}^*_{\text{BD}}$
    \item Compute a targeted free energy estimate $\Delta \hat{F}_{\text{targ}}^{0\rightarrow1}$ using the mapping with parameters  $\boldsymbol{\theta}^*_{\text{BD}}$ using BAR (Eq. \ref{eq:bar})
    \item Train for $n_{\text{epochs, HD}}$ using $\mathcal{L}_{\text{HD}}(\boldsymbol{\theta}, \Delta \hat{F}_{\text{targ}}^{0\rightarrow1})$ (Eq. \ref{eq:hd_loss}) to obtain final mapping parameters $\boldsymbol{\theta}^*$. 
\end{enumerate}
This procedure takes advantage of the theoretically favorable behavior of the BD loss at very low overlap, before switching to the HD loss which is directly related to the BAR variance. In this work we use $n_{\text{epochs, BD}}=n_{\text{epochs, HD}}=n_{\text{epochs}}/2$. We refer to this as the BD-HD training strategy for MNFs. 

\section{Models and Numerical Methodology}
\label{methodology}
\subsection*{General Mapping Evaluation Methodology}

Generating training and evaluation data for our mappings is comprised of three main steps --- simulate states 0 and 1 in the LAMMPS simulation package \cite{RN139} to form the training set (collecting $N_\text{ind}$ independent configurations from each state), create $N_\text{rep}$ sets of independent replicates of the training simulations of states 0 and 1 to form the evaluation set, and finally create a set of simulations applying a windowing method between states 0 and 1 to get a high accuracy ground truth free energy difference. 

All free energy calculations (applying the BAR estimator) are conducted using the PyMBAR Python package \cite{RN45}. Independence between simulated configurations is determined by examining the energy autocorrelation function \cite{RN136}. Our evaluation sets provide access to $N_\text{rep}$ independent targeted free energy estimates, and we consolidate this information by reporting a ``typical" targeted free energy estimate for the number of samples considered. This is given by the average free energy prediction over the evaluation set $\left(\Delta\hat{F}^{0\rightarrow1}=\left<\Delta\hat{F}^{0\rightarrow1}_i\right>_{i\in\text{rep}}\right)$, with the uncertainty given as the square root of the average variance over the replicates $\left(\sigma(\Delta\hat{F}^{0\rightarrow1})=\sqrt{\left<\sigma^2(\Delta\hat{F}^{0\rightarrow1})\right>_{i\in\text{rep}}}\right)$. It should be noted that while Eq. \ref{eq:bar_var} is only formally valid asymptotically ($N_\text{ind}\rightarrow\infty$), and thus we compute errors in free energies for each evaluation replicate using the bootstrapping method provided by the PyMBAR package, with 100 bootstraps \cite{RN45}. We corroborate our variance calculation approach by ensuring the agreement of our computed accelerations with those calculated by taking the standard deviation of the free energy over the evaluation set replicas. In order to report accelerations from our method conservatively, for large accelerations at lower overlap (where BAR finite sample error estimates can be noisy), we report $\tilde{n}$ in approximate orders of magnitude rather than providing exact numbers. It is important to note that evaluating trained mappings on an independent evaluation set has been shown to be important in dealing with pathologies such as bias that appear when targeted free energies are only evaluated on the training set \cite{RN190,RN191}.

In all test cases considered, the potential energy between a pair of atoms can be described as a sum of non-coulombic (Lennard-Jones) and coulombic (point charge) interactions --- 
\begin{equation}
\label{eq:pot_eng}
    u_{jk}(r_{jk})=\underbrace{4\epsilon_{jk}\left[\left(\frac{\sigma_{jk}}{r_{jk}}\right)^{12}-\left(\frac{\sigma_{jk}}{r_{jk}}\right)^6\right]}_{u_{jk}^\text{LJ}}+\underbrace{\frac{q_jq_k}{4\pi\epsilon_0r_{jk}}}_{u_{jk}^\text{coul}}
\end{equation}
where $r_{jk}$ is the distance between atoms, $\epsilon_{jk}$ and $\sigma_{jk}$ are pair LJ parameters, $q_j$ is the charge of atom $j$ ($q_j=0$ for all $j$ in the LJ test cases), and $\epsilon_0$ is the permittivity of free space. The total potential energy of our systems can be expressed as --- 
\begin{equation}
     U_{i,j}(\lambda_i,\phi_j) =  \sum_{k \ne \text{ins}} \sum_{t >k} \left[u^{\text{LJ}}_{kt} + u^{\text{coul}}_{kt}\right] + \sum_{k \ne \text{ins}} \left[\lambda_i u^{\text{LJ}}_{k,\text{ins}}+\phi_j u^{\text{coul}}_{k,\text{ins}}\right]
    \label{eq:U_ion_pair}
\end{equation}
where $\lambda_i\in[0,1]$ and $\phi_j\in[0,1]$ are coupling parameters that control the interaction of our inserted particle with the remainder of the system. For our free energy estimation challenges of interest states 0 and 1 are typically taken to be two pairs of coupling parameter values.

\subsection*{Pure LJ Fluid}
\label{sec:pure_lj}
The first test case of the effectiveness of our methodology is a pure LJ fluid under \emph{NVT} conditions. The interaction between two particles (indexed $j$ and $k$) is given by the first term in Eq. \ref{eq:pot_eng}, since our particles are uncharged. Since we consider a pure fluid, our parameters $\epsilon_{jk}=\epsilon$ and $\sigma_{jk}=\sigma$ are the same for all pairs. We set our $\epsilon$ and $\sigma$ parameters to those of pure Argon ($\epsilon/k_B=93.3\text{ K}$, $\sigma=3.542\text{ \AA}$), and specify our system conditions by setting the values of the reduced density ($\rho^*=\sigma^3\rho$) and reduced temperature ($T^*=k_B T/\epsilon)$ \cite{RN152}. 

We estimate the free energy difference of inserting an extra Argon particle (given index 1) into a system. Our system potential energy is given by Eq. \ref{eq:U_ion_pair}, with $q=0$ and thus $u^{\text{coul}}=0$. State 0 is our ``not inserted" state ($\lambda_0=0.0$) and state 1 is our ``inserted" state ($\lambda_1=1.0$). This is exactly the system Hahn and Then considered, where we are attempting to accelerate the estimation of the chemical potential $(\mu^{0\rightarrow1})$ \cite{RN152}. Here we train a radial mapping that moves solvent particles uniformly towards/away from the inserted solute, as defined by Eq. \ref{eq:radial_map}.

The systems we consider have 215 solvent particles and a single solute ($N=216$). To train the 1D MNF for this system, we run MD simulations of states 0 and 1 in the LAMMPS simulation package \cite{RN139}, and train our mapping by minimizing our losses represented as ensemble averages over independent configurations generated from these simulations.  Detailed simulation procedures are provided in the SI. The effectiveness of our mapping trained with a given loss is evaluated as described previously, with all simulations comprising the training and evaluation sets providing $N_\text{ind}=3334$ independent samples from each state.

\subsection*{Binary LJ Mixture}
\label{sec:mixture_lj}

To test the applicability of our radial mapping scheme to multi-component systems, we also examine a binary LJ mixture that exhibits solid-solution phase behavior. This solution is made up of ``A" (solute) and ``B" (solvent) LJ beads ($q=0$ for all beads and thus the  pair energy is given by the first term of Eq. \ref{eq:pot_eng}), where we are attempting to accelerate the calculation of the solute chemical potential ($\mu_A$). The LJ parameters used in this system are those examined by Espinosa \emph{et al.} \cite{RN21} and are provided in the SI. 

As a generalization of the pure fluid case, we now train different 1D radial mappings for each species present. If our system has $N_A$ solute beads, $N_B$ solvent beads, and a single ``inserted" solute particle ($N=N_A+N_B+1$), a configuration can be represented as $\left\{\textbf{r}_{\text{ins}},\textbf{r}^{(A)}_1, \dots,\textbf{r}^{(A)}_{N_A},\textbf{r}^{(B)}_1, \dots,\textbf{r}^{(B)}_{N_B}\right\}$. Thus, we transform our configurations according to the expression -
\begin{equation}
\begin{aligned}
    \phi\left(\textbf{r}_{\text{ins}},\textbf{r}^{(A)}_1, \dots,\textbf{r}^{(A)}_{N_A},\textbf{r}^{(B)}_1, \dots,\textbf{r}^{(B)}_{N_B}\right)=\left\{\textbf{r}_{\text{ins}},f^{(A)}\left(\textbf{r}^{(A)}_1,\textbf{r}_{\text{ins}}\right), \dots,f^{(B)}\left(\textbf{r}^{(B)}_{N_B},\textbf{r}_{\text{ins}}\right)\right\}
\end{aligned}   
\end{equation}
where $f^{(i)}$ is defined analogously to Eq. \ref{eq:radial_map}, but has a radial mapping function $\psi^{(i)}$ with different parameters trained for each possible kind of bead of type $i\in\{A,B\}$ that the inserted solute particle can interact with. 

The specific system we consider is comprised of $N_B=500$ solvent particles and $N_A=15$ solute particles along with the single solute (``A") particle we are inserting ($N=516$). This corresponds to a solute mole fraction of $x_A\approx0.03$ which is just below the solubility limit for this system reported by Espinosa \emph{et al}. \cite{RN21} We again consider state 0 as the inserted solute behaving as an ideal gas ($\lambda_0=0.0$ in Eq. \ref{eq:U_ion_pair}) and state 1 as the solute fully interacting ($\lambda_1=1.0$ in Eq. \ref{eq:U_ion_pair}). Exact simulation specifications are given in Sec. 1 of the SI. 

Similar to the pure case, we train our mapping using configurations from one set of simulations of states 0 and 1, with $N_\text{ind} = 3334$ configurations from each state. However, our evaluation procedure is slightly different --- we again have 5 independent evaluation sets, however each set has 4 different sets of simulations of states 0 and 1, each run for a different number of simulation steps ($n_{\text{steps}}/10^6=\{20, 50, 100, 500\}$). This allows us to gauge mapping performance as a function of the number of independent samples collected. 

\subsection*{Dilute Electrolyte Solution}
\label{sec:electrolyte}

The last test case we consider is a technologically relevant system --- the hydration behavior of an ion in a system of water molecules. In this case, we examine a single $\text{Na}^+$ or $\text{Cl}^-$ ion (described by the Joung-Cheatham (JC) force field) \cite{RN38} in a box of $N_w=216$ water molecules described by the SPC/E force field (a rigid, 3-site water model) \cite{RN212}. Uniform charge backgrounds are present in the systems as needed to maintain overall charge neutrality at all instances. The interaction between two atoms is given by the pair potential as defined in Eq. \ref{eq:pot_eng}. FF parameters for these systems are provided in the SI. 

We test the ability of our method to accelerate free energy calculations in two distinct ion hydration scenarios --- 
\begin{itemize}
    \item Free Energy of Charging (for both $\text{Na}^+$ and $\text{Cl}^-$): State 0 corresponds to $\lambda_0=1.0, \phi_0=0.0$ and state 1 corresponds to $\lambda_1=1.0, \phi_1=1.0$, as defined in Eq. \ref{eq:U_ion_pair}.
    \item FF Parameter Perturbation for $\text{Na}^+$: State 0 corresponds to $\lambda_0=1.0, \phi_0=1.0$ and state 1 corresponds to $\lambda_1=1.0, \phi_1=1.0$, as defined in Eq. \ref{eq:U_ion_pair}, except in state 1 we now have $\epsilon_{\text{Na}^+} = 0.4297 \text{ kcal/mol}$ (the epsilon parameter of $\text{K}^+$ in the JC FF \cite{RN38}), $\sigma_{\text{Na}^+} = 2.5 \text{ \AA}$ (a substantial increase in ion size), and $q_{\text{Na}^+} = 0.85 \text{ e}$ (a common charge value employed in scaled-charge electrolyte FFs \cite{RN100}), compared to the unmodified sodium ion FF parameters given in the SI, with the typical mixing rules maintained. 
\end{itemize}

In these systems, we also apply a radial mapping as defined in Eq. \ref{eq:radial_map} to the ion-O distances, however this case differs from the LJ systems in that our solvent is now anisotropic. Thus, in addition to ion-O distances being significantly affected by the insertion, another difference is the orientation of water molecules as a function of their radial distance from the ion. As the charge of the inserted ion increases, water molecules reorient to point towards/away from the ion, depending on the sign of the ion charge. This motivates the use of a two-dimensional MNF for the electrolyte test case, and a schematic of such a mapping and its behavior is provided in the right panel of Fig. \ref{fig:flow_demo}. We construct our 2D MNF to consist of two steps. First, the mapping rigidly pushes water molecules away from the inserted ion while maintaining their orientation --- this is essentially an application of our 1D MNF construction to the ion-oxygen distance. Second, it rotates the water molecule as to modify the angle between the ion-oxygen vector and the H-O-H bisector of the water molecule --- we refer to this angle as $\gamma$. The magnitude of this rotation depends on the radial position of the water molecule being mapped. In practice, this 2D mapping is summarized by the operation $\left(r,\cos\gamma\right)\rightarrow\left(r',\cos\gamma'\right)$, and this is shown visually in the right panel of Fig. \ref{fig:flow_demo}. This transformation has a simple analytical expression for its Jacobian that can be computed efficiently and is provided in the SI. 

\subsection*{MNF Training}

Training our MNFs with any $f$-divergence requires recalculating the forward and reverse mapped works (given by Eqs. \ref{eq:wf_mapped} and \ref{eq:wr_mapped}) and their gradients at each training step, as these are the quantities modified by our mapping parameters. This is by far the largest computational bottleneck in the training process. To accelerate this, we introduce a physically informed assumption --- that our radial mappings only need to deviate from identity up to some short-range cutoff radius, which we refer to as a ``fitting limit" $r_f$. Thus, our radial and orientational mapping functions are now defined as --- 
\begin{equation}
    \psi^{(SR)}(r) = \begin{cases}
    \psi(r) & \text{if } r < r_f \\
    r   & \text{if } r \geq r_f
\end{cases}
\end{equation}
\begin{equation}
    g^{(SR)}(\gamma|r) = \begin{cases}
    g(\gamma|r) & \text{if } r < r_f \\
    \gamma   & \text{if } r \geq r_f
\end{cases}
\end{equation}
Here we enforce that $\left(\psi^{(SR)}\right)'(r_f)=1$. This means that for our 1D or 2D mappings, only atoms/molecules within a sphere of radius $r_f$ from the inserted solute have their positions modified. 

This assumption allows us to accelerate our MNF training, because as mentioned before, the biggest computational challenge is calculating a quantity such as $U_1(\phi(x,\boldsymbol{\theta}))$ at every training step. But, given our ``short-range" mapping formulation, we can re-write the above quantity as --- 
\begin{equation}
\begin{aligned}
\label{eq:U_in_out}
        U_1(\phi(x,\boldsymbol{\theta})) = \sum_{i\in \text{I}} \sum_{ \text{ }j>i} u^{(1)}_{ij}(\phi(\textbf{r}_i,\textbf{r}_j,\textbf{r}_{\text{ins}},\boldsymbol{\theta}))+\sum_{i\in \text{E}} \sum_{j\in \text{E}, \text{ }j>i} u^{(1)}_{ij}(r_{ij})
\end{aligned}
\end{equation}
here indices refer individual atoms or molecules, and $I$ refers to the group of atoms/molecules within the fitting limit $r_f$ from the inserted particle (``Interior" group, assume they have the first $N_I$ indices), $E$ refers to the group of molecules outside of this inner sphere (``Exterior" group), and $u^{(1)}_{ij}$ refers to the interaction energy in state 1 (where $x$ is sampled from state 0). The key observation here is that only the first sum depends on our mapping parameters $\boldsymbol{\theta}$, and thus, we can pre-compute $U_1(x)$ (and equivalently $U_0(y)$), and then only compute the quantity $\Delta U^{(t)}_1 = U_1(\phi(x,\boldsymbol{\theta}_t)) -U_1(x)$ at each training step $t$, where the second term in Eq. \ref{eq:U_in_out} cancels upon evaluating this difference. This calculation has algorithmic complexity $\mathcal{O}(N_I\cdot N)$ for short-range interactions and $\mathcal{O}(N_I\cdot N_k)$ for long-range interactions (Ewald sums), where $N_I$ is the number of atoms inside the fitting limit, $N_k$ is the number of $k$-vectors used in the Ewald sum, and $N$ is the total number of atoms in the system. Explicit expressions for $\Delta U^{(t)}_1$ for the pair potentials we consider are given in the SI. 

It should be noted that $r_f$ is a tunable parameter, and it needs to be large enough such that the trained mapping is not truncated artificially in a region where it would otherwise be not equal to the identity map. A simple way to determine $r_f$ is to start with a large value and find at what radial distance the trained mapping relaxes to the identity map. We find conservative values of $r_f$ to be between $6-7 \text{ \AA}$ and demonstrate in the SI (through examples of trained mappings) that our trained mappings relax to the identity map well within this radius. 

All automatic differentiation and vectorization operations needed to efficiently implement our MNF training procedures in Python are provided by the JAX package.\cite{jax2018github} To implement our mapping functions $\psi$ and $g$, we use rational quadratic splines as proposed by Durkan \emph{et al}. \cite{RN199}. This framework ensures that all mapping functions are differentiable and bijective, and we construct our MNFs such that all symmetries present in a molecular simulation box are obeyed. The details of our implementation of such splines are provided in Sec. 6 of the SI. 

\section{Results}
\label{results}

\subsection*{Chemical Potential of Pure LJ Fluid}

We first examine the ability of our method to find 1D radial maps that accelerate the calculation of the chemical potential of a pure LJ fluid. We consider such a fluid (with specifications given in Sec. \ref{sec:pure_lj} and the SI) at conditions $\rho^*=0.9$ and $T^*=1.2$, as this was the test system considered by Hahn \& Then \cite{RN152}. This system showcases the effectiveness of our approach, as our mapping trained with the BD-HD training strategy provides a $\tilde{n}\sim10^1-10^2$ (approximately two orders of magnitude) times acceleration of the chemical potential calculation relative to the identity mapping, and a $\tilde{n}\sim10^1$ (approximately one order of magnitude) times acceleration relative to the KLD-trained mapping. 

\begin{figure}[h!]
\centering
\includegraphics[width=8 cm]{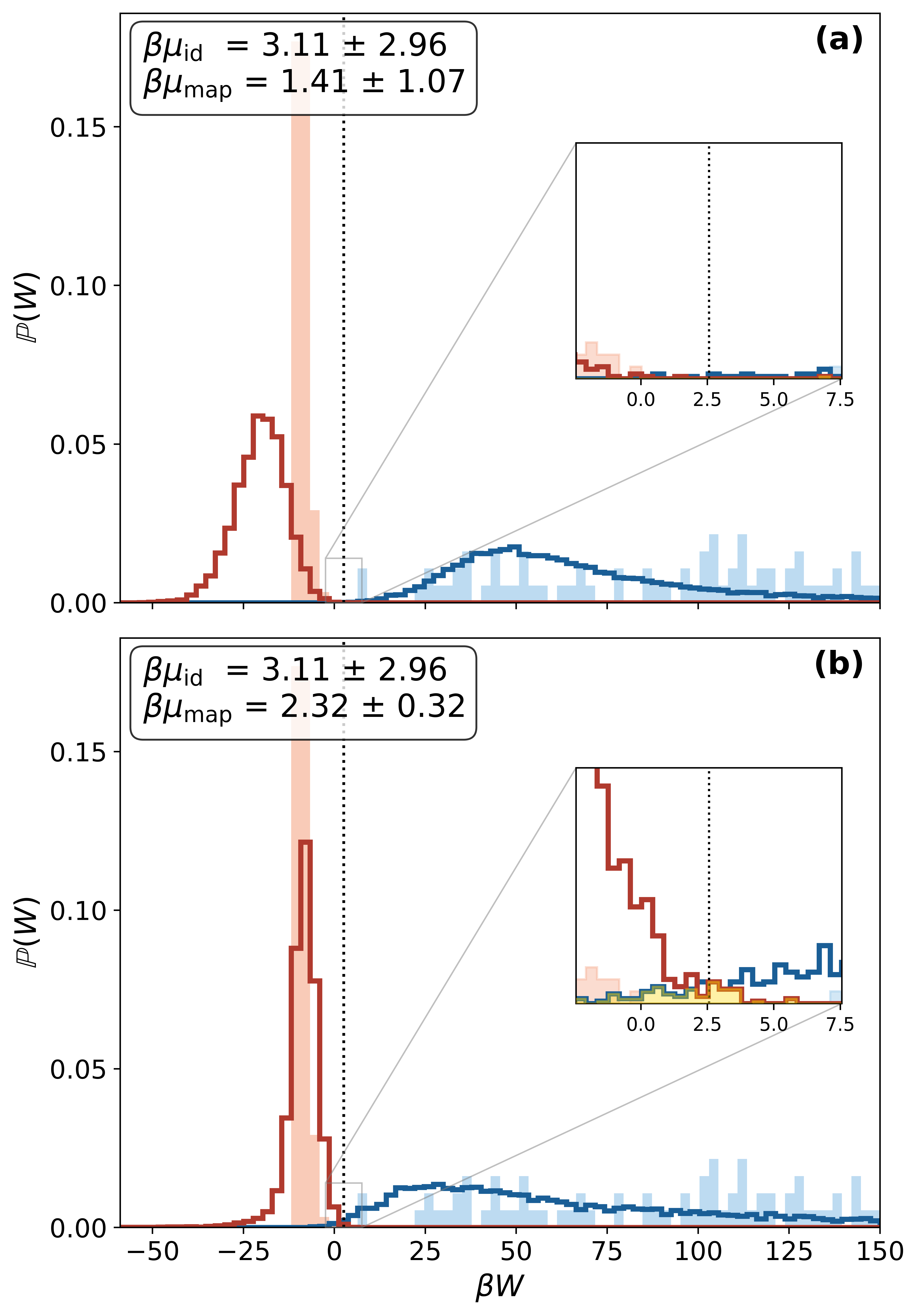}
\caption{\textbf{Mapped Work Distributions for the Chemical Potential of a Pure LJ Fluid}. Both panels show the forward (blue) and reverse (orange) work distributions under the identity mapping, the ground truth chemical potential value ($\beta\mu$, vertical dotted line), and an inset magnifying the overlap region ($\beta W =\beta\mu\pm5$) with overlap shaded in yellow. (a) Forward (blue line) and Reverse (red line) mapped work distributions after $n_\text{epochs}=1000$ epochs of training with the KLD loss given in Eq. \ref{eq:loss_kl} (b) Forward (blue line) and Reverse (red line) mapped work distributions after $n_\text{epochs}=1000$ epochs of training with the BD-HD training procedure (500 epochs per stage).}
\label{fig:pure_lj_mapped_work}
\end{figure}

From our results, it is clear how the choice of loss function affects the performance of our resulting 1D MNF. Fig. \ref{fig:pure_lj_mapped_work} shows the resulting work distributions after training our MNF with a KLD loss function (top panel), and after training with our BD-HD strategy (bottom panel). 
The shown work distributions adhere to the behavior of the CFT where the probabilities of mapped and unmapped work distributions are equal exactly at the chemical potential of our pure fluid. As we theorized, training with the KLD minimizes the distance between the means of the mapped work distributions. In this case this is accomplished by sacrificing the proximity of $-\left<\tilde{W}^{(R)}(y,\boldsymbol{\theta})\right>_1$ to $\Delta F$, while largely minimizing the distance of $\left<\tilde{W}^{(F)}(x,\boldsymbol{\theta})\right>_0$ to $\Delta F$, settling at a Pareto optimal solution for such a restricted mapping. The pathology we noted in Sec. \ref{sec:kld_pathology} is explicitly shown in the inset of the top panel of Fig. \ref{fig:pure_lj_mapped_work} --- even though the distances between the means are minimized, the relevant overlap in region $W\approx\Delta F$ is not meaningfully enhanced, as shown by the modest reduction in the standard error of our targeted chemical potential estimate ($\beta\hat{\mu}_{\text{KL}}^{0\rightarrow1}=1\pm1$) relative to the identity mapping ($\beta\hat{\mu}^{0\rightarrow1}=3\pm3$). On the other hand, training with the BD-HD loss (shown in the lower panel of Fig. \ref{fig:pure_lj_mapped_work}) targets overlap and achieves the maximum increase in overlap possible within our restricted mapping class (overlap is shown as the shaded yellow region under the probability histograms). This is expressed in the approximately ten times lower standard error in our targeted chemical potential ($\beta\hat{\mu}_{\text{BD-HD}}^{0\rightarrow1}=2.3\pm0.3$) relative to the identity mapping. It should also be noted that this targeted chemical potential estimate is within statistical uncertainty of the ground truth value, as we computed the ground truth chemical potential (using 10, equally spaced $\lambda$ windows) to be $\beta\mu^{0\rightarrow1}=2.56\pm0.04$.

We compare our method of mapping construction to the RDF-matching heuristic of Hahn \& Then \cite{RN152}.  The best performing mapping constructed from their intuitive / ``heuristic-based" approaches had a converged Harmonic overlap value of $\tilde{\mathcal{H}}_{0,1}=1.9\times10^{-3}$, while our BD-HD trained mapping achieves a converged Harmonic overlap value of $\tilde{\mathcal{H}}_{0,1}=7.0\times10^{-3}$ \cite{RN152}. This corresponds to a $\sim3.7$ times increase in the Harmonic overlap using our methodology which, according to Eq. \ref{eq:bar_var}, corresponds to about a 50\% reduction in standard error of our chemical potential estimate. We are confident that we have arrived at the (essentially) optimal 1D radial mapping for this system. 

It is also important to note the computational expense of mapping generation. In this case, we achieve a training speed of 0.03 s/epoch on a single GPU, which amounts to about 30 seconds of GPU time to achieve the accelerations we see above. All training is conducted on a NVIDIA A100 GPU, with 80 GB of GPU memory.

\subsection*{Chemical Potential of Binary LJ Mixture}

\begin{figure}[h!]
\centering
\includegraphics[width=8 cm]{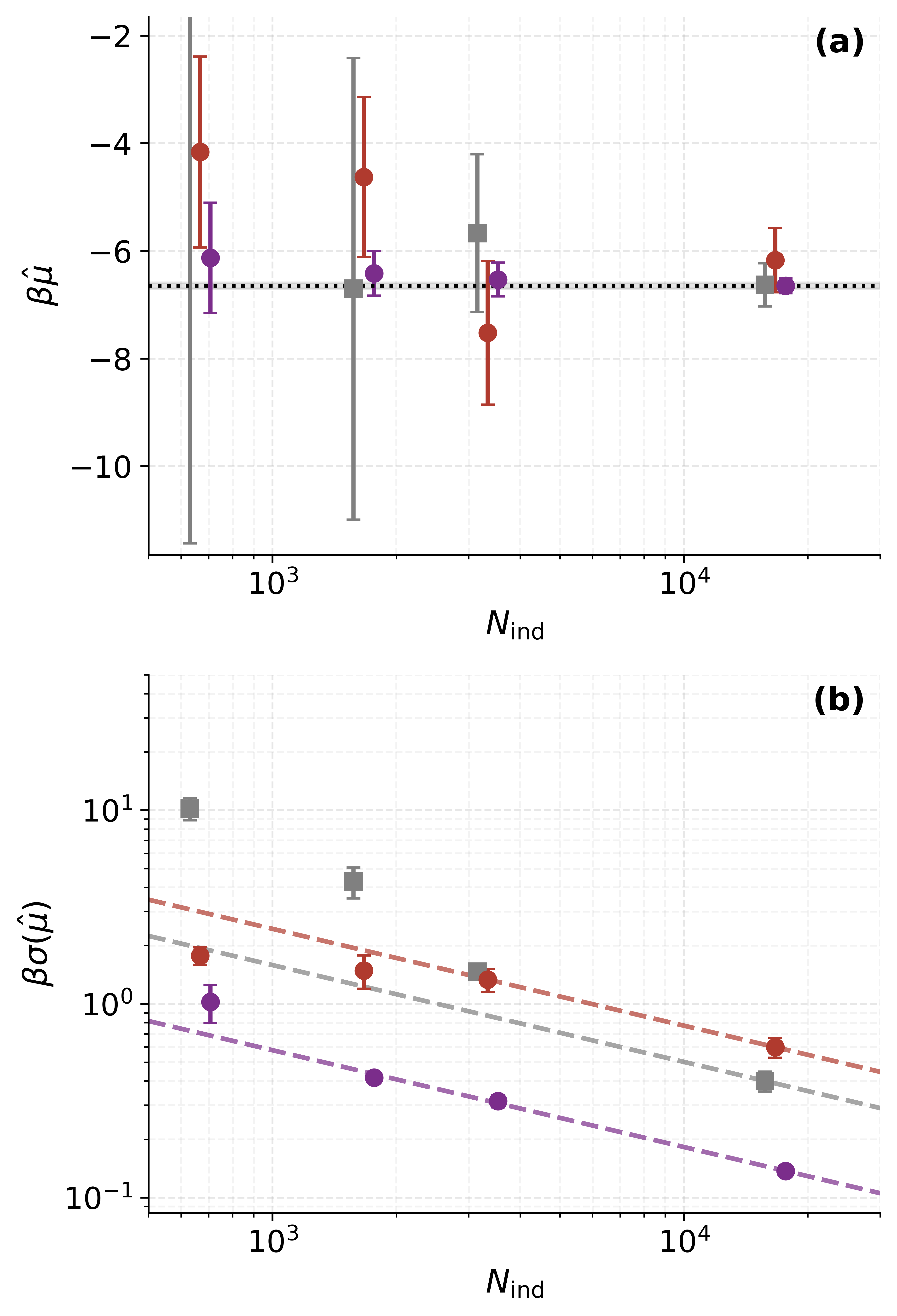}
\caption{\textbf{Acceleration of the Chemical Potential Calculation for a Binary LJ System}. (a) Targeted free energy estimates ($\beta\hat{\mu}$) as a function of the number of independent samples collected from each state ($N_{\text{ind}}$) for the identity mapping (grey), and mappings trained with the KLD (red) and BD-HD (purple) training approaches. (b) Standard error in the targeted free energy estimates ($\beta\sigma(\hat{\mu})$) as a function of the number of independent samples collected from each state ($N_{\text{ind}}$) for the identity mapping (grey), and mappings trained with the KLD (red) and BD-HD (purple) training approaches. Dashed lines represent $\beta\sigma(\hat{\mu})=K\cdot N_{\text{ind}}^{-1/2}$, where K is a constant such that the guideline intersects the final point of each data series.}
\label{fig:mix_lj_scaling}
\end{figure}

We next examine the ability of our 1D mapping scheme to handle multi-component systems. As described in Sec. \ref{sec:mixture_lj} we compute (and accelerate) the calculation of the solute (A) chemical potential in a binary LJ mixture (of A and B particles), where here we define different radial mappings acting on the distance between the inserted solute and particles of type A or B. Fig. \ref{fig:mix_lj_scaling} shows our results for applying the identity mapping and mappings trained with the KLD and BD-HD loss functions to this system, and the scaling of these results with the number of independent samples collected. We see that the effectiveness of the BD-HD training strategy extends to multi-component systems, providing about a ten-fold acceleration of the calculation of the solute chemical potential relative to the identity map.

Interestingly, we can see that on the evaluation set for this system, the KLD-trained MNF provides worse performance than both the identity mapping and the BD-HD trained MNF. This provides an even more extreme example of the KLD pathology identified in Sec. \ref{sec:kld_pathology}, where training with the KLD loss motivates the sacrifice of existing probability mass in the $W\approx\Delta F$ region in order to minimize the distance between the means of the forward and reverse mapped work distributions. The upper panel of Fig. \ref{fig:mix_lj_scaling} shows how the BD-HD estimate (purple point) retains low bias (relative to the ground truth value of $\beta\Delta F = -6.65\pm0.06$) and error even in very low data regimes where $N_{\text{ind}}<10^3$. On the other hand, this free energy estimation problem is entirely intractable for the identity map in such low-data regimes, as can be seen by the huge error bars on its chemical potential estimate. 

The lower panel of Fig. \ref{fig:mix_lj_scaling} shows the dependence of the standard error in the (targeted) chemical potential estimate as a function of the number of independent samples collected for the identity mapping, as well as our two differently-trained MNFs. The points and error bars (68 \% confidence intervals) are generated by taking the square root of the average variance over the 5 evaluation replicates and the standard error of the standard errors of the 5 evaluation replicates for each simulation length. The dashed lines show the regime of $\sigma(\hat{\mu})\propto N^{-1/2}$, with each line anchored to the last point of each data series. This regime is approximately achieved for all estimators for $N_{\text{ind}}>5\times10^3$, but we can see that the targeted estimate from the BD-HD trained MNF achieves this regime for all numbers of independent samples collected. As $N_{\text{ind}}\rightarrow\infty$, we can see that the BD-HD trained estimate achieves an asymptotic acceleration of about $\tilde{n}\approx10$ relative to the identity map, with this acceleration of course being greater in low-data regimes ($N_{\text{ind}}<3\times10^3$). These results, along with the fact that mapping training for this system is approximately as fast as the pure LJ case, confirm that the 1D MNFs we propose (together with our novel training strategy), can effectively accelerate chemical potential calculations in both single, and multi-component systems. 

\subsection*{Hydration of Single Ions}

Finally, we test our approach on a challenging, technologically relevant system that  has not previously been treated by targeted free energy estimates --- the hydration of alkali/halide ions in water. We specifically focus on the insertion of a single sodium ion ($\text{Na}^+$) or a single chloride ion ($\text{Cl}^-$) into a box of pure water as described in Sec. \ref{sec:electrolyte}. The most challenging part of chemical potential calculations involving ions is the charging step --- where ionic charges are slowly turned on. This is equivalent to varying the coupling parameter $\phi$ from 0 to 1, with the system potential energy defined by Eq. \ref{eq:U_ion_pair}. We focus specifically on accelerating the calculation of this charging free energy ($\Delta G^{\text{coul}}$) by training and applying our 2D MNF. Since this calculation cannot practically be done in 1 step, we examine the performance of our mapping as function of the number of windows (intermediate two-state free energy calculations) used to compute $\Delta G^{\text{coul}}$, where $N_{\text{win}}=1.0/\Delta\phi$ and $\Delta\phi$ is the window spacing (e.g. $\Delta\phi=0.5$ would correspond to 2 windows, where we first compute the free energy between states $\phi=0.0$ and $\phi=0.5$ and then $\phi=0.5$ and $\phi=1.0$). Thus, we compute the total free charging free energy and its corresponding standard error as ---  
$\Delta G^{\text{coul}}=\sum_{i=1}^{N_\text{win}}\Delta G^{\text{coul}}_{i,i+1}$ and 
$\sigma\left(\Delta G^{\text{coul}}\right)=\sqrt{\sum_{i=1}^{N_\text{win}}\sigma^2_{i,i+1}}$, where $\Delta G^{\text{coul}}_{i,i+1}$ is the free energy difference between the state $i$ with coupling parameter value $\phi_i$ and $i+1$ with coupling parameter value $\phi_{i+1}$, and $\sigma^2_{i,i+1}$ is the corresponding per-sample variance in the estimate. 

\begin{figure}[h!]
\centering
\includegraphics[width=8 cm]{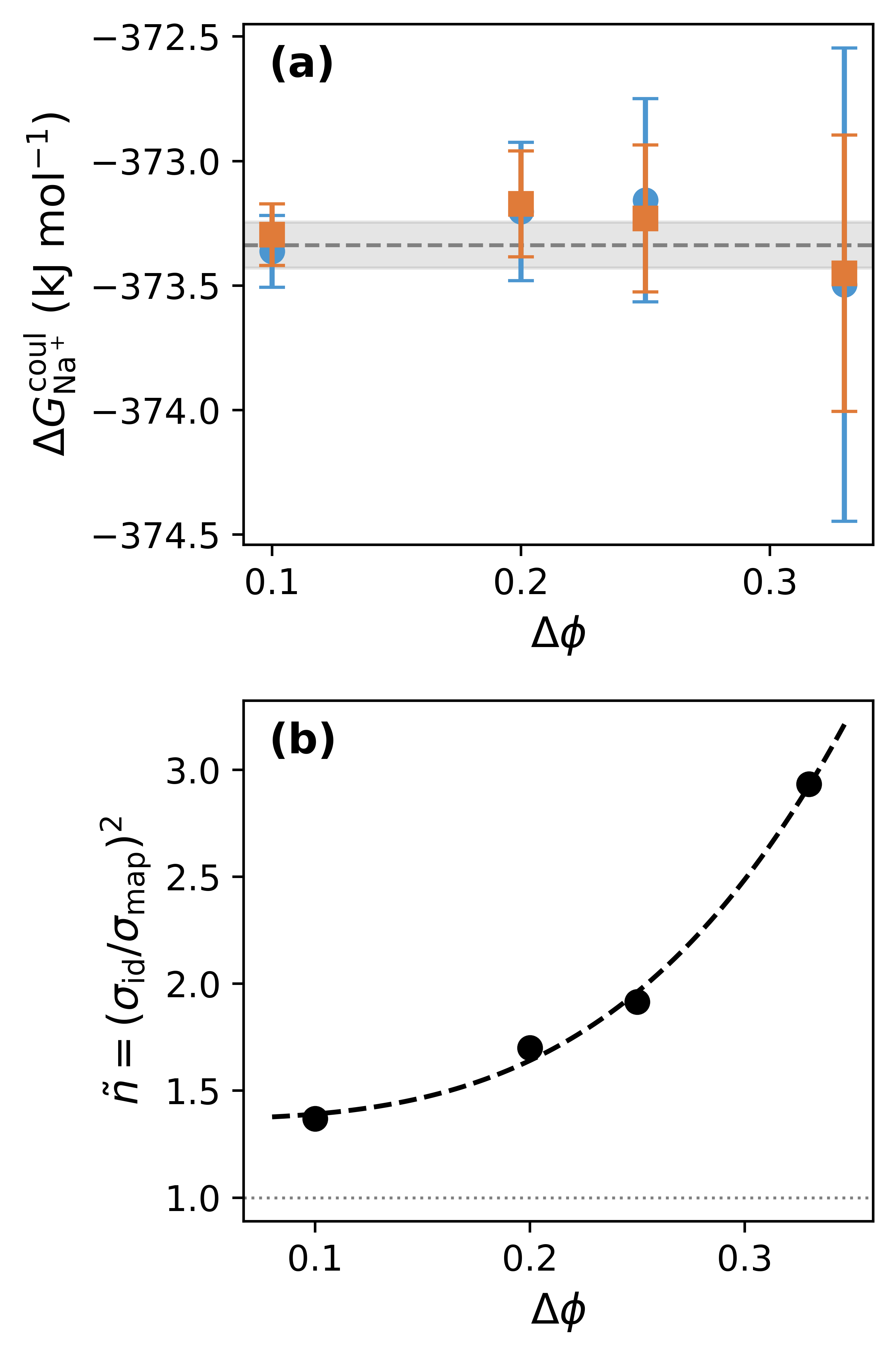}
\caption{\textbf{Acceleration of the Charging Free Energy Calculation for $\text{Na}^+$}. (a) Charging free energy ($\Delta G^{\text{coul}}$) versus window size ($\Delta\phi$) for the identity mapping (blue) and the targeted estimate (orange) trained using the BD-HD procedure. (b) Effective sample size increase ($\tilde{n}$) versus window size ($\Delta\phi$) for the BD-HD mapping compared to the identity mapping. The dashed line represents a power law fit to the data to guide the eye.}
\label{fig:electrolyte_dphi}
\end{figure}

Fig. \ref{fig:electrolyte_dphi} shows our results for applying our 2D MNF trained using the BD-HD strategy to the charging of $\text{Na}^+$. The top panel shows absolute charging free energy estimates using the identity mapping (blue) and our trained MNF (orange) as a function of the coupling parameter window width. The targeted estimate achieves consistently lower standard error on the evaluation dataset as compared to the identity mapping. The bottom panel quantifies exactly how much of an acceleration our targeted estimate provides. We can see that the greatest acceleration is seen for the coarsest windowing ($\Delta\phi=1/3$, $\tilde{n}\approx3$), while only slight gains are seen for the highest windowing resolution ($\Delta\phi=1/10$, $\tilde{n}\approx1.4$). An equivalent plot showing results for the charging free energy of $\text{Cl}^-$ is provided in Sec. 10 of the SI. Generally, our 2D MNF still provides lower error at each window width, but provides a smaller acceleration of the calculation relative to the $\text{Na}^+$ case. For the $\text{Cl}^-$ case, we see an acceleration of $\tilde{n}\approx1.6$ at $\Delta\phi=1/3$ and $\tilde{n}\approx1.2$ at $\Delta\phi=1/10$. 

It is important to note that calculating the chemical potential of single ions is subject to significant finite size effects \cite{RN23}. This is due to the uniform, neutralizing background charge density, introduced to non-neutral simulation boxes to maintain charge neutrality, that the system interacts with. It has been previously reported that this system size effect can be removed by computing the chemical potential at a given concentration for various system sizes and extrapolating the single-ion chemical potential to the thermodynamic limit \cite{RN15, RN23}. Thus, we show that our 2D MNFs trained to accelerate the insertion of $\text{Na}^+$ into 216 waters, can be applied to a larger system at the same concentration (648 waters and 1 ion pair) without re-training. For $\Delta\phi=1/4$, the mapping trained on the 216 water molecule system provided an equivalent $\tilde{n}\approx2$ acceleration on the charging free energy calculation for the 648 water system, showing that acceleration results are transferable to larger system sizes without retraining, and a figure showing these results is provided in Sec. 11 of the SI. 

\begin{figure}[h!]
\centering
\includegraphics[width=10 cm]{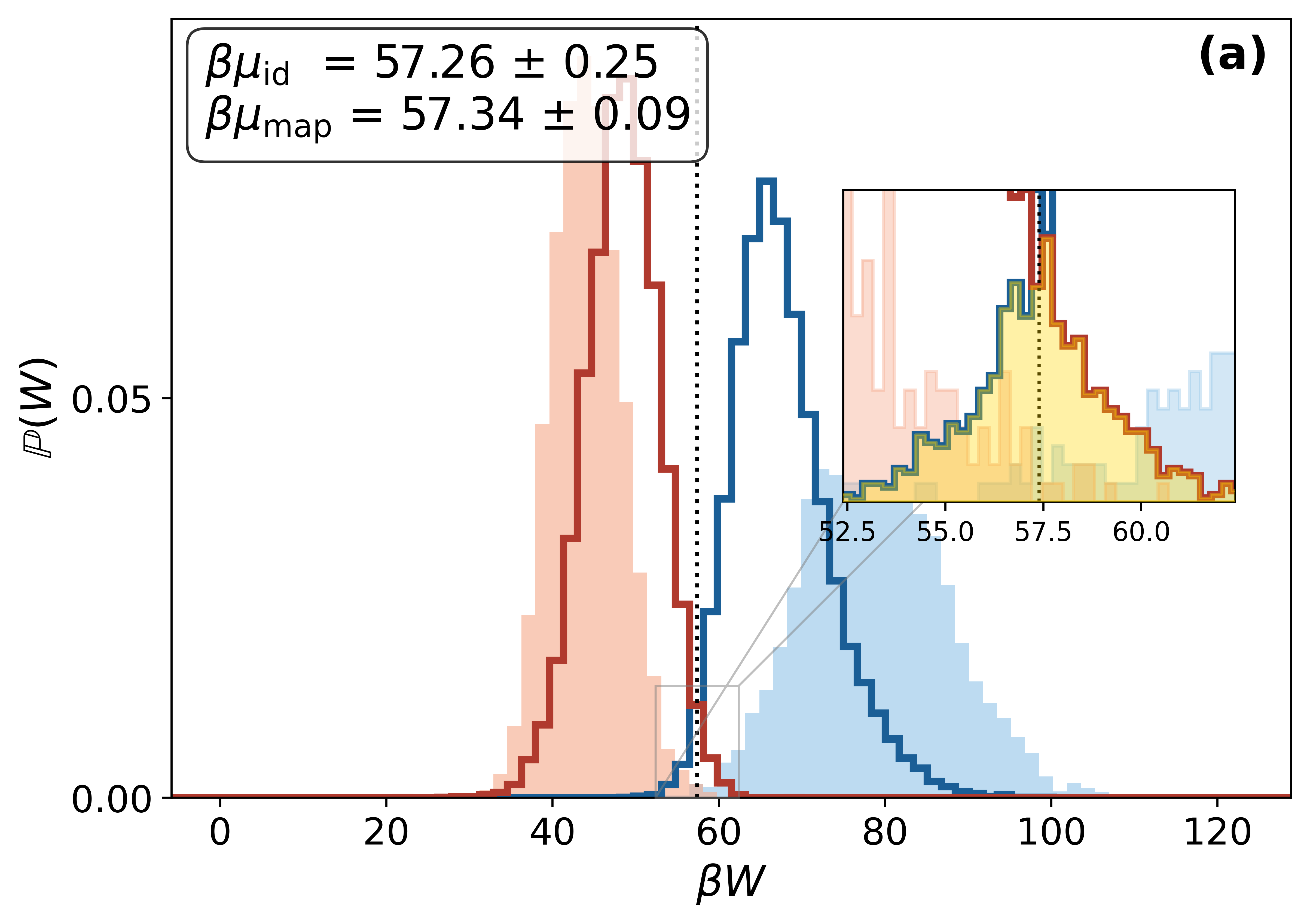}
\caption{\textbf{Mapped Work Distributions for the Free Energy Change of Adjusting the $\text{Na}^+$ FF}. Forward (light blue) and reverse (light orange) work distributions under the identity mapping are shown. The ground truth chemical potential value ($\beta\mu$) is given by the vertical dotted line, and an inset magnifying the overlap region ($\beta W =\beta\mu\pm5$) is displayed with overlap shaded in yellow. The forward and reverse mapped work distributions after $n_\text{epochs}=1000$ epochs of training with the BD-HD training strategy are shown by the solid blue and solid red lines respectively.}
\label{fig:electrolyte_perturb}
\end{figure}

The free energy change of inserting a single ion into a pure water box is used to assess hydration free energy predictions of different force fields. Thus, it is useful to accelerate calculations of free energy differences between systems with different sets of ion FF parameters. As described in Sec. \ref{sec:electrolyte}, we start with a fully interacting sodium ion in solution using the SPC/E+JC FF (with the exact FF parameters provided in Sec. 1 of the SI), and we perturb the ion parameters to $\epsilon_{\text{Na}^+} = 0.4297 \text{ kcal/mol}$, $\sigma_{\text{Na}^+} = 2.5 \text{ \AA}$ , and $q_{\text{Na}^+} = 0.85 \text{ e}$. We then train the same 2D MNF architecture as the charging case, using the BD-HD procedure, to accelerate the calculation of the free energy difference between these states. Fig. \ref{fig:electrolyte_perturb} shows this targeted free energy estimate, where our mapping significantly increases configuration space overlap, leading to a $\tilde{n}\approx8$ acceleration in the free energy estimate relative to the identity mapping. This significant acceleration suggests the 2D MNF is particularly well-suited to capturing the structural changes associated with this FF parameter perturbation. Our identity and mapped estimates agree with our high accuracy ground truth value (computed with 10 windows) of $\beta\mu=57.39\pm0.02$. 

Finally, we note that all 2D MNFs for electrolyte systems are trained at a speed of about $0.06-0.07\text{ s/epoch}$, which amounts to about a minute for 1000 epochs of training on a single GPU, about twice as slow as the LJ cases. 

\section{Discussion and Conclusions}
\label{discussion}

In this work, we have shown that training 1D and 2D MNFs with the BD-HD strategy can accelerate free energy calculations in diverse solvation scenarios while requiring at most minutes of GPU time for training. Below we discuss the strengths, weaknesses, and remaining questions regarding our approach. 

A significant advantage of our method is its low computational cost compared to the actual MD simulations used to generate independent samples from states 0 and 1.  Training for all cases examined takes 0.5--1.5 minutes on a single GPU, while simulations typically take tens of CPU hours (especially for electrolyte systems). Current BGs with many parameters can take days to train on multiple GPUs, so our training costs are 3-4 orders of magnitude lower \cite{RN188}. Of course, these more expressive architectures provide greater accelerations of free energy calculations, but, in our opinion, the speed of our method mitigates any limitations in its performance. MNFs can be trained and immediately applied at negligible cost for any new chemical potential calculation in any new system. Due to its efficiency, our approach can be viewed as a way to go beyond the provably minimum variance a BAR estimate provides, as the training cost is negligible compared to the cost of simulating states 0 and 1. 

Despite the negligible cost, we were able to accelerate free energy calculations involving the solvation of single ions in water, a system that, to our knowledge, has not yet been accelerated with generative modeling methods due to its complexity as a molecular liquid with long-range coulombic interactions. Our framework is also trivially transferable between system sizes with no additional training cost, possessing a feature that is not present in typical BGs without significant additional architectural complexity \cite{schebek2025scalable,RN193}. This shows the advantage that the simple, physically-informed nature of MNFs provides.

In our theoretical discussions, we consider three potential losses for MNF training --- KLD, BD, and HD. Based on our theoretical arguments about how BD is independent of $\Delta F^{0\rightarrow1}$ and has more robust gradient behavior at low overlap, while HD is directly related to the BAR variance, we settled on our hybrid BD-HD strategy for training MNFs. While we have showed the failure of the KLD as a training loss, what about simply training on just BD or HD? A detailed comparison of the behavior of all 4 training strategies is presented in Sec. 8 of the SI. There, we examine a slightly more challenging pure LJ fluid at conditions $\rho^*=1.1$ and $T^*=2.5$. We use this test case to confirm our theoretical results that the HD loss has very few samples determining its gradient at low overlap (potentially exacerbating overfitting to noise in the training set), and that its minimum can be affected by a biased initial estimate of $\Delta \hat{F}^{0\rightarrow1}$. However, it is worth noting that for the MNFs considered in this work, with very low numbers of parameters ($\sim 10^2-10^3$), training with BD, HD, and BD-HD losses provides very similar results for almost all cases. Even in pathological cases such as the one presented in the SI, BD-HD training gives similar results to simply training with BD and only provides a small standard error reduction on the training set compared to training with only HD, and free energy estimates on the evaluation set are essentially identical. Similarly, in the SI we numerically show the more robust gradient behavior of the BD loss (compared to the HD loss) at low overlap, but this does not lead to significant differences in optimization results. We believe that this is due to the simplicity of the optimization landscape for these simple test cases, and these issues will likely be exaggerated if more complex MNFs are considered, but this is beyond the scope of this work. We also examined a second test case for our loss comparisons --- the charging of $\text{Na}^+$ in water from $\phi=0.0$ to $\phi=0.5$. Training strategy comparisons on the training and evaluation sets mirror the LJ results discussed above, but in this case, it turns out that all four losses do equally well --- the KLD training does not give significantly worse results in this case. For charging free energy calculations, the forward and reverse work distributions are often symmetrical (similar to each other in shape, with different means), which is different from the stark forward-reverse asymmetry of particle insertion scenarios. The KLD turns out to still be effective in this case, as minimizing the distance between the work distribution means happens to align with maximizing overlap. Despite the KLD training being non-pathological here, in general it is not possible to know \emph{a priori} when minimizing the distance between the mapped work distribution means will reliably increase phase space overlap between essentially disjoint distributions mapped using a non-expressive flow. 

Given these empirical results, and the fact that training with all four loss strategies is of equivalent expense (apart from the added cost of a single BAR evaluation in the middle of training for BD-HD, $\sim5-10$ s), we believe that our recommendation practically collapses to training MNFs with either the BD loss or the BD-HD hybrid strategy, as this approach will reliably find the (approximately) most effective mapping achievable for a given MNF architecture. Also, the KLD training strategy should be avoided when training MNFs, especially for particle insertion free energy calculations, as we have demonstrated its poor performance for such systems. While theoretically sound, the practical effects of our distinctions between the BD and HD losses are largely conjectural, and should be explored in future work with potentially more complex MNFs. 

It is rather interesting to see the variability in performance of our mappings for different systems. Our 1D MNF significantly accelerated chemical potential calculations in the pure LJ system and the binary LJ mixture. Our 2D MNF significantly accelerated our calculation of the free energy change upon FF parameter perturbation, but provided only modest gains for the charging of $\text{Na}^+$ in water, and hardly accelerated the calculation of the charging free energy of $\text{Cl}^-$ at all. This reveals the main limitation of our methodology --- the structure of the transformation the MNF applies is still intuition-based, and thus it is impossible to know how effective the free energy acceleration will be \emph{a priori}. The major difference between MNFs and previous, intuition-based approaches is that here, for a given intuitive transformation, we are confident that we are finding the (approximately) most effective transformation of that type. For example, we showed that we were able to achieve a 3.7-fold overlap improvement for the case of the pure LJ chemical potential compared to the heuristic approach of Hahn \& Then \cite{RN152}. Despite this deficiency, we argue that the low computational cost of our approach justifies its application, as we have not observed our MNFs under-performing the identity mapping for any of the cases tested, while adding negligible computational cost.

However, this begs the question --- how can MNFs be extended? Where do we derive more complex, and potentially more effective, low dimensional transformations, perhaps even for situations beyond solvation? One option, at least for solvation scenarios is to climb the Kirkwood hierarchy --- while our current 1D MNF affects pair coordinates relative to the solute, we also can design MNFs that modify triplet coordinates $\left(r,r',\theta\right)$ relative to the solute \cite{RN135}. In addition, it may be interesting to apply dimensionality reduction techniques such as autoencoders to find dominant features of liquid state configurations \cite{RN211}, perhaps motivating new generalized directions for MNFs to act on. However, we believe that our ideas apply to any non-fully expressive NF. Perhaps it would be interesting to start with a maximally expressive flow, begin taking away coupling layers, and examine to what extent our training strategy can still produce meaningful accelerations. Overall, we believe that the MNF framework provides future avenues to fully take advantage of Jarzynski's seminal result by better balancing the training cost and transferability of mappings with their effectiveness in enhancing free energy estimates, all simply by modifying our training strategy and baking in physical intuition. 

In summary, we have presented a novel approach to efficiently constructing bijective mappings that can be used for targeted free energy estimation. Our ``minimal" normalizing flows are low-dimensional, physically informed mappings that are non-expressive by construction. However, they can be trained in about one minute on a single GPU to meaningfully enhance the phase space overlap between two states. This is compared to the typical GPU days it takes to train typical, fully expressive flows. Our method is built upon our observation that the KLD fails as a loss function for non-expressive flows, and we propose an alternative loss strategy that directly targets the enhancement of phase space overlap. This loss strategy (the BD-HD approach) is based on the combination of the Bhattacharyya distance with the harmonic overlap metric, which mitigates theoretical issues with training directly on the harmonic overlap metric when the configuration spaces of the two Boltzmann distributions in question have negligible overlap. We show how our mappings can provide accelerations to the calculations of the chemical potential in pure ($\tilde{n}\sim10^1-10^2$) and binary (asymptotic $\tilde{n}\sim10^1$) LJ systems, as well as calculations of the free energy change upon ion charging in water ($\tilde{n}\approx3$ at $\Delta\phi=0.33$) and FF perturbation of $\text{Na}^+$ in water ($\tilde{n}\approx8$), all while retaining training speeds of less than 0.1 seconds/epoch. In addition, due to how our MNFs are constructed, they are immediately transferable across system sizes, enabling rapid training in small systems and applications to larger ones, as we demonstrate for the case of the charging free energy of $\text{Na}^+$. This is very important as calculations of both single-ion and ion pair chemical potentials depend significantly on system size, and thus require extrapolations to the thermodynamic limit \cite{RN15,RN23}. Our method has the limitation that it is impossible to know \emph{a priori} to what degree the physically-informed transformations our MNFs act on can increase the phase space overlap between our states of interest. This is seen in how variable the accelerations we obtain are for the various systems we test --- for some test cases, our MNFs act on the dominant variance-reducing channels, while in others they seem to largely miss them. Thus, future work should focus on finding low-dimensional transformations that better guarantee the enhancement of phase space overlap while not being fully expressive and thus retaining low training cost. This will likely lead to an increase in MNF complexity, but walking the tightrope of training efficiency vs. acceleration in the convergence of free energy estimates in industrially relevant systems is an exciting direction for future work. We believe this work lays the foundation for developing physically-informed, generative-model-assisted free energy estimates that can efficiently enhance our ability to assess phase behavior in complex systems. 

\newpage  

\section*{Supplementary Material}
The supplementary material accompanying this article contains LAMMPS simulation details for all simulation data we generated (Sec. 1), a detailed derivation of the correlation function expression in Eq. \ref{eq:cf_decomp} (Sec. 2), derivations showing the pathologies of the HD loss (Sec. 3), mathematical considerations for neglecting SBDs (Sec. 4), a derivation of the analytically expressible Jacobian of our 2D MNF (Sec. 5), details of the spline implementations for our 1D and 2D MNFs and a description of the optimization hyper-parameters we use during training (Sec. 6), detailed expressions for how we accelerate MNF training using the fitting limit assumption (Sec. 7), numerical comparisons between all possible loss strategies we consider (Sec. 8), examples of what our trained 1D and 2D MNFs look like (Sec. 9), results for the acceleration of the charging free energy of $\text{Cl}^-$ (Sec. 10), and finally, results for the system size transferability of our 2D MNFs (Sec. 11).

\section*{Acknowledgments}
Financial support for this work was provided by the Office of Basic Energy Sciences, U.S. Department of Energy, under Award DESC0002128. P.B.B. acknowledges support from the National Science Foundation Graduate Research Fellowship Program under Grant No. DGE-2039656. Any opinions, findings, and conclusions or recommendations expressed in this material are those of the authors and do not necessarily reflect the views of the National Science Foundation. We would like to thank Ryan Szukalo for useful discussions and feedback on the manuscript. 

\section*{Author Declarations}
\subsection*{Conflict of Interest}
The authors have no conflicts to declare.

\section*{Data Availability}

Computer codes used in this work, including scripts to set up and run MD simulations for the pure LJ, binary LJ, and ion hydration test cases, and to implement the methods described in Secs. \ref{background} \& \ref{methodology} to efficiently train both 1D and 2D MNFs for all of the test systems described are freely available for download from the Princeton Data Commons repository at \href{https://doi.org/10.34770/nrpq-qk33}{https://doi.org/10.34770/nrpq-qk33}.

% \printbibliography
\bibliography{references}

@software{jax2018github,
  author = {James Bradbury and Roy Frostig and Peter Hawkins and Matthew James Johnson and Yash Katariya and Chris Leary and Dougal Maclaurin and George Necula and Adam Paszke and Jake Vander{P}las and Skye Wanderman-{M}ilne and Qiao Zhang},
  title = {{JAX}: composable transformations of {P}ython+{N}um{P}y programs},
  url = {http://github.com/jax-ml/jax},
  version = {0.3.13},
  year = {2018},
}

@article{RN212,
   author = {Berendsen, H. J. C. and Grigera, J. R. and Straatsma, T. P.},
   title = {The missing term in effective pair potentials},
   journal = {J. Phys. Chem.},
   volume = {91},
   number = {24},
   pages = {6269-6271},
   DOI = {10.1021/j100308a038},
   year = {1987},
   type = {Journal Article}
}

@article {MR10358,
    AUTHOR = {Bhattacharyya, A.},
     TITLE = {On a measure of divergence between two statistical populations
              defined by their probability distributions},
   JOURNAL = {Bull. Calcutta Math. Soc.},
  FJOURNAL = {Bulletin of the Calcutta Mathematical Society},
    VOLUME = {35},
      YEAR = {1943},
     PAGES = {99--109},
      ISSN = {0008-0659},
}

@book{McQuarrie2000,
  author    = {McQuarrie, D. A.},
  title     = {Statistical Mechanics},
  publisher = {University Science Books},
  address   = {Sausalito, CA},
  year      = {2000},
  isbn      = {978-1-891389-15-3}
}

@article{RN211,
   author = {Chen, W and Ferguson, AL},
   title = {Molecular Enhanced Sampling with Autoencoders: On-The-Fly Collective Variable Discovery and Accelerated Free Energy Landscape Exploration},
   journal = {J. Comput. Chem.},
   volume = {39},
   number = {25},
   pages = {2079-2102},DOI = {10.1002/jcc.25520},
   year = {2018},
   type = {Journal Article}
}

@article{RN210,
   author = {Shirts, MR and Bair, E and Hooker, G and Pande, VS},
   title = {Equilibrium free energies from nonequilibrium measurements using maximum-likelihood methods},
   journal = {Phys. Rev. Lett.},
   volume = {91},
   number = {14},DOI = {10.1103/PhysRevLett.91.140601},
   year = {2003},
   type = {Journal Article}
}

@article{RN209,
  author   = {Baron, P. B. and Panagiotopoulos, A. Z.},
  title    = {Efficient calculation of crystal-solution coexistence lines for aqueous electrolytes},
  journal  = {J. Chem. Phys.},
  volume   = {163},
  number   = {22},
  year     = {2025},
  doi      = {10.1063/5.0304180}
}

@article{RN206,
  author   = {Crooks, G. E.},
  title    = {Path-ensemble averages in systems driven far from equilibrium},
  journal  = {Phys. Rev. E},
  volume   = {61},
  number   = {3},
  pages    = {2361--2366},
  year     = {2000},
  doi      = {10.1103/PhysRevE.61.2361}
}

@article{RN203,
  author   = {Dellago, C. and Hummer, G.},
  title    = {Computing Equilibrium Free Energies Using Non-Equilibrium Molecular Dynamics},
  journal  = {Entropy},
  volume   = {16},
  number   = {1},
  pages    = {41--61},
  year     = {2014},
  doi      = {10.3390/e16010041}
}

@article{RN205,
  author   = {Jarzynski, C.},
  title    = {Nonequilibrium equality for free energy differences},
  journal  = {Phys. Rev. Lett.},
  volume   = {78},
  number   = {14},
  pages    = {2690--2693},
  year     = {1997},
  doi      = {10.1103/PhysRevLett.78.2690}
}

@article{RN208,
  author   = {Papamakarios, G. and Nalisnick, E. and Rezende, D. J. and Mohamed, S. and Lakshminarayanan, B.},
  title    = {Normalizing Flows for Probabilistic Modeling and Inference},
  journal  = {J. Mach. Learn. Res.},
  volume   = {22},
  year     = {2021}
}

@article{RN207,
  author   = {Pham, T. T. and Shirts, M. R.},
  title    = {Optimal pairwise and non-pairwise alchemical pathways for free energy calculations of molecular transformation in solution phase},
  journal  = {J. Chem. Phys.},
  volume   = {136},
  number   = {12},
  year     = {2012},
  doi      = {10.1063/1.3697833}
}

@article{RN204,
  author   = {Zhong, A. and Kuznets-Speck, B. and DeWeese, M. R.},
  title    = {Time-asymmetric fluctuation theorem and efficient free-energy estimation},
  journal  = {Phys. Rev. E},
  volume   = {110},
  number   = {3},
  year     = {2024},
  doi      = {10.1103/PhysRevE.110.034121}
}

@article{RN200,
  author   = {Aykol, M. and Kim, S. and Hegde, V. I. and Snydacker, D. and Lu, Z. and Hao, S. Q. and Kirklin, S. and Morgan, D. and Wolverton, C.},
  title    = {High-throughput computational design of cathode coatings for Li-ion batteries},
  journal  = {Nat. Commun.},
  volume   = {7},
  year     = {2016},
  doi      = {10.1038/ncomms13779}
}

@inproceedings{RN199,
  author      = {Durkan, C. and Bekasov, A. and Murray, I. and Papamakarios, G.},
  title       = {Neural Spline Flows},
  booktitle   = {Advances in Neural Information Processing Systems},
  editor      = {Wallach, H. M. and Larochelle, H. and Beygelzimer, A. and d'Alch{\'e}-Buc, F. and Fox, E. A. and Garnett, R.},
  publisher   = {Curran Associates, Inc.},
  year        = {2019},
  volume      = {32},
  pages       = {7511--7522},
  url         = {https://papers.nips.cc/paper/8969-neural-spline-flows}
}

@article{RN197,
  author   = {Ghidini, A. and Serra, E. and Cavalli, A.},
  title    = {On Free Energy Calculations in Drug Discovery},
  journal  = {Acc. Chem. Res.},
  volume   = {58},
  number   = {20},
  pages    = {3137--3145},
  year     = {2025},
  doi      = {10.1021/acs.accounts.5c00465}
}

@article{RN202,
  author   = {Panagiotopoulos, A. Z.},
  title    = {Monte Carlo methods for phase equilibria of fluids},
  journal  = {J. Phys. Condens. Matter},
  volume   = {12},
  number   = {3},
  pages    = {R25-R52},
  year     = {2000},
  doi      = {10.1088/0953-8984/12/3/201}
}

@article{RN201,
  author   = {Sadybekov, A. V. and Katritch, V.},
  title    = {Computational approaches streamlining drug discovery},
  journal  = {Nature},
  volume   = {616},
  number   = {7958},
  pages    = {673--685},
  year     = {2023},
  doi      = {10.1038/s41586-023-05905-z}
}

@article{RN198,
  author   = {Vega, C. and Sanz, E. and Abascal, J. L. F. and Noya, E. G.},
  title    = {Determination of phase diagrams via computer simulation: methodology and applications to water, electrolytes and proteins},
  journal  = {J. Phys. Condens. Matter},
  volume   = {20},
  number   = {15},
  year     = {2008},
  doi      = {10.1088/0953-8984/20/15/153101}
}

@article{RN194,
  author   = {Ahmad, R. and Cai, W.},
  title    = {Free energy calculation of crystalline solids using normalizing flows},
  journal  = {Modell. Simul. Mater. Sci. Eng.},
  volume   = {30},
  number   = {6},
  year     = {2022},
  doi      = {10.1088/1361-651X/ac7f4b}
}

@inproceedings{RN196,
  author      = {Klein, L. and Kr{\"a}mer, A. and No{\'e}, F.},
  title       = {Equivariant Flow Matching},
  booktitle   = {Advances in Neural Information Processing Systems},
  editor      = {Oh, A. and Naumann, T. and Globerson, A. and Saenko, K. and Hardt, M. and Levine, S.},
  publisher   = {Curran Associates, Inc.},
  year        = {2023},
  volume      = {36},
  url         = {https://proceedings.neurips.cc/paper_files/paper/2023/hash/bc827452450356f9f558f4e4568d553b-Abstract-Conference.html}
}

@article{RN195,
  author   = {Wirnsberger, P. and Papamakarios, G. and Ibarz, B. and Racanière, S. and Ballard, A. J. and Pritzel, A. and Blundell, C.},
  title    = {Normalizing flows for atomic solids},
  journal  = {Mach. Learn.: Sci. Technol.},
  volume   = {3},
  number   = {2},
  year     = {2022},
  doi      = {10.1088/2632-2153/ac6b16}
}

@article{Noe2019boltzmann,
  author   = {No{\'e}, F. and Olsson, S. and K{\"o}hler, J. and Wu, H.},
  title    = {Boltzmann generators: Sampling equilibrium states of many-body systems with deep learning},
  journal  = {Science},
  volume   = {365},
  number   = {6457},
  pages    = {eaaw1147},
  year     = {2019},
  doi      = {10.1126/science.aaw1147}
}

@online{Schebek2025solvation,
  author      = {Schebek, M. and Frob{\"o}se, N. M. and Keller, B. G. and Rogal, J.},
  title       = {Estimating solvation free energies with {B}oltzmann generators},
  date        = {2025-12},
  eprinttype  = {arxiv},
  eprint      = {2512.18147},
  eprintclass = {cond-mat.stat-mech},
  url         = {https://arxiv.org/abs/2512.18147},
  doi         = {10.48550/arXiv.2512.18147}
}

@inproceedings{Midgley2023flow,
  author      = {Midgley, L. I. and Stimper, V. and Simm, G. N. C. and Sch{\"o}lkopf, B. and Hern{\'a}ndez-Lobato, J. M.},
  title       = {Flow Annealed Importance Sampling Bootstrap},
  booktitle   = {Proceedings of the 11th International Conference on Learning Representations},
  publisher   = {OpenReview.net},
  year        = {2023},
  eprinttype  = {arxiv},
  eprint      = {2208.01893},
  url         = {https://arxiv.org/abs/2208.01893}
}

@online{Felardos2023designing,
  author      = {Felardos, L. and H{\'e}nin, J. and Charpiat, G.},
  title       = {Designing losses for data-free training of normalizing flows on {B}oltzmann distributions},
  date        = {2023-01},
  eprinttype  = {arxiv},
  eprint      = {2301.05475},
  eprintclass = {cs.LG},
  url         = {https://arxiv.org/abs/2301.05475},
  doi         = {10.48550/arXiv.2301.05475}
}

@online{schebek2025scalable,
  author      = {Schebek, M. and No{\'e}, F. and Rogal, J.},
  title       = {Scalable {B}oltzmann generators for equilibrium sampling of large-scale materials},
  date        = {2025-09},
  eprinttype  = {arxiv},
  eprint      = {2509.25486},
  eprintclass = {cond-mat.stat-mech},
  url         = {https://arxiv.org/abs/2509.25486},
  doi         = {10.48550/arXiv.2509.25486}
}

@article{RN191,
  author   = {Olehnovics, E. and Liu, Y. M. and Mehio, N. and Sheikh, A. Y. and Shirts, M. R. and Salvalaglio, M.},
  title    = {Assessing the Accuracy and Efficiency of Free Energy Differences Obtained from Reweighted Flow-Based Probabilistic Generative Models},
  journal  = {J. Chem. Theory Comput.},
  volume   = {20},
  number   = {14},
  pages    = {5913--5922},
  year     = {2024},
  doi      = {10.1021/acs.jctc.4c00520}
}

@article{RN192,
  author   = {Olehnovics, E. and Liu, Y. M. and Mehio, N. and Sheikh, A. Y. and Shirts, M. R. and Salvalaglio, M.},
  title    = {Accurate Lattice Free Energies of Packing Polymorphs from Probabilistic Generative Models},
  journal  = {J. Chem. Theory Comput.},
  volume   = {21},
  number   = {5},
  pages    = {2244--2255},
  year     = {2025},
  doi      = {10.1021/acs.jctc.4c01612}
}

@article{RN185,
  author   = {Paliwal, H. and Shirts, M. R.},
  title    = {Multistate reweighting and configuration mapping together accelerate the efficiency of thermodynamic calculations as a function of molecular geometry by orders of magnitude},
  journal  = {J. Chem. Phys.},
  volume   = {138},
  number   = {15},
  year     = {2013},
  doi      = {10.1063/1.4801332}
}

@article{RN190,
  author   = {Rizzi, A. and Carloni, P. and Parrinello, M.},
  title    = {Targeted Free Energy Perturbation Revisited: Accurate Free Energies from Mapped Reference Potentials},
  journal  = {J. Phys. Chem. Lett.},
  volume   = {12},
  number   = {39},
  pages    = {9449--9454},
  year     = {2021},
  doi      = {10.1021/acs.jpclett.1c02135}
}

@article{RN188,
  author   = {Schebek, M. and He, J. J. and Hoffmann, E. and Du, Y. Q. and Noé, F. and Rogal, J.},
  title    = {Assessing generative modeling approaches for free energy estimates in condensed matter},
  journal  = {J. Chem. Phys.},
  volume   = {164},
  number   = {18},
  year     = {2026},
  doi      = {10.1063/5.0320214}
}

@article{RN184,
  author   = {Schieber, N. P. and Shirts, M. R.},
  title    = {Configurational mapping significantly increases the efficiency of solid-solid phase coexistence calculations via molecular dynamics: Determining the FCC-HCP coexistence line of Lennard-Jones particles},
  journal  = {J. Chem. Phys.},
  volume   = {150},
  number   = {16},
  year     = {2019},
  doi      = {10.1063/1.5080431}
}

@inproceedings{RN193,
  author      = {Tan, C. B. and Bose, A. J. and Lin, C. and Klein, L. and Bronstein, M. M. and Tong, A.},
  title       = {Scalable Equilibrium Sampling with Sequential {Boltzmann} Generators},
  booktitle   = {Proceedings of the 42nd International Conference on Machine Learning},
  editor      = {Singh, A. and Fazel, M. and Hsu, D. and Lacoste-Julien, S. and Berkenkamp, F. and Maharaj, T. and Wagstaff, K. and Zhu, J.},
  publisher   = {PMLR},
  year        = {2025},
  volume      = {267},
  pages       = {58467--58498},
  url         = {https://proceedings.mlr.press/v267/tan25a.html}
}

@article{RN186,
  author   = {Tan, T. B. and Schultz, A. J. and Kofke, D. A.},
  title    = {Efficient calculation of temperature dependence of solid-phase free energies by overlap sampling coupled with harmonically targeted perturbation},
  journal  = {J. Chem. Phys.},
  volume   = {133},
  number   = {13},
  year     = {2010},
  doi      = {10.1063/1.3483899}
}

@article{RN187,
  author   = {Willow, S. Y. and Kang, L. L. and Minh, D. D. L.},
  title    = {Learned mappings for targeted free energy perturbation between peptide conformations},
  journal  = {J. Chem. Phys.},
  volume   = {159},
  number   = {12},
  year     = {2023},
  doi      = {10.1063/5.0164662}
}

@article{RN183,
  author   = {Coretti, A. and Falkner, S. and Geissler, P. L. and Dellago, C.},
  title    = {Learning mappings between equilibrium states of liquid systems using normalizing flows},
  journal  = {J. Chem. Phys.},
  volume   = {162},
  number   = {18},
  year     = {2025},
  doi      = {10.1063/5.0253034}
}

@article{RN152,
  author   = {Hahn, A. M. and Then, H.},
  title    = {Using bijective maps to improve free-energy estimates},
  journal  = {Phys. Rev. E},
  volume   = {79},
  number   = {1},
  year     = {2009},
  doi      = {10.1103/PhysRevE.79.011113}
}

@article{RN151,
  author   = {Jarzynski, C.},
  title    = {Targeted free energy perturbation},
  journal  = {Phys. Rev. E},
  volume   = {65},
  number   = {4},
  year     = {2002},
  doi      = {10.1103/PhysRevE.65.046122}
}

@article{RN153,
  author   = {Schebek, M. and Invernizzi, M. and Noé, F. and Rogal, J.},
  title    = {Efficient mapping of phase diagrams with conditional Boltzmann Generators},
  journal  = {Mach. Learn.: Sci. Technol.},
  volume   = {5},
  number   = {4},
  year     = {2024},
  doi      = {10.1088/2632-2153/ad849d}
}

@article{RN182,
  author   = {Wirnsberger, P. and Ballard, A. J. and Papamakarios, G. and Abercrombie, S. and Racanière, S. and Pritzel, A. and Rezende, D. J. and Blundell, C.},
  title    = {Targeted free energy estimation via learned mappings},
  journal  = {J. Chem. Phys.},
  volume   = {153},
  number   = {14},
  year     = {2020},
  doi      = {10.1063/5.0018903}
}

@article{RN159,
  author   = {Baron, P. B. and Panagiotopoulos, A. Z.},
  title    = {Top-down optimization of aqueous electrolyte force fields to model chemical potentials and solubilities},
  journal  = {J. Chem. Phys.},
  volume   = {162},
  number   = {21},
  year     = {2025},
  doi      = {10.1063/5.0272640}
}

@article{RN166,
  author   = {Benavides, A. L. and Aragones, J. L. and Vega, C.},
  title    = {Consensus on the solubility of NaCl in water from computer simulations using the chemical potential route},
  journal  = {J. Chem. Phys.},
  volume   = {144},
  number   = {12},
  year     = {2016},
  doi      = {10.1063/1.4943780}
}

@article{RN100,
  author   = {Benavides, A. L. and Portillo, M. A. and Chamorro, V. C. and Espinosa, J. R. and Abascal, J. L. F. and Vega, C.},
  title    = {A potential model for sodium chloride solutions based on the TIP4P/2005 water model},
  journal  = {J. Chem. Phys.},
  volume   = {147},
  number   = {10},
  year     = {2017},
  doi      = {10.1063/1.5001190}
}

@article{RN18,
  author   = {Bennett, C. H.},
  title    = {Efficient estimation of free energy differences from {Monte Carlo} data},
  journal  = {J. Comput. Phys.},
  volume   = {22},
  number   = {2},
  pages    = {245--268},
  year     = {1976},
  doi      = {10.1016/0021-9991(76)90078-4}
}

@article{RN136,
  author   = {Chodera, J. D. and Swope, W. C. and Pitera, J. W. and Seok, C. and Dill, K. A.},
  title    = {Use of the weighted histogram analysis method for the analysis of simulated and parallel tempering simulations},
  journal  = {J. Chem. Theory Comput.},
  volume   = {3},
  number   = {1},
  pages    = {26--41},
  year     = {2007},
  doi      = {10.1021/ct0502864}
}

@article{RN6,
  author   = {Dockal, J. and Lisal, M. and Moucka, F.},
  title    = {Molecular Force Field Development for Aqueous Electrolytes: 2. Polarizable Models Incorporating Crystalline Chemical Potential and Their Accurate Simulations of Halite, Hydrohalite, Aqueous Solutions of NaCl, and Solubility},
  journal  = {J. Chem. Theory Comput.},
  volume   = {16},
  number   = {6},
  pages    = {3677--3688},
  year     = {2020},
  doi      = {10.1021/acs.jctc.0c00161}
}

@article{RN21,
  author   = {Espinosa, J. R. and Young, J. M. and Jiang, H. and Gupta, D. and Vega, C. and Sanz, E. and Debenedetti, P. G. and Panagiotopoulos, A. Z.},
  title    = {On the calculation of solubilities via direct coexistence simulations: Investigation of NaCl aqueous solutions and Lennard-Jones binary mixtures},
  journal  = {J. Chem. Phys.},
  volume   = {145},
  number   = {15},
  year     = {2016},
  doi      = {10.1063/1.4964725}
}

@article{RN38,
  author   = {Joung, I. S. and Cheatham, T. E.},
  title    = {Determination of alkali and halide monovalent ion parameters for use in explicitly solvated biomolecular simulations},
  journal  = {J. Phys. Chem. B},
  volume   = {112},
  number   = {30},
  pages    = {9020--9041},
  year     = {2008},
  doi      = {10.1021/jp8001614}
}

@article{RN135,
  author   = {Kirkwood, J. G.},
  title    = {Statistical mechanics of fluid mixtures},
  journal  = {J. Chem. Phys.},
  volume   = {3},
  number   = {5},
  pages    = {300--313},
  year     = {1935}
}

@article{RN7,
  author   = {Panagiotopoulos, A. Z.},
  title    = {Simulations of activities, solubilities, transport properties, and nucleation rates for aqueous electrolyte solutions},
  journal  = {J. Chem. Phys.},
  volume   = {153},
  number   = {1},
  year     = {2020},
  doi      = {10.1063/5.0012102}
}

@article{RN23,
  author   = {Saravi, S. H. and Panagiotopoulos, A. Z.},
  title    = {Individual Ion Activity Coefficients in Aqueous Electrolytes from Explicit-Water Molecular Dynamics Simulations},
  journal  = {J. Phys. Chem. B},
  volume   = {125},
  number   = {30},
  pages    = {8511--8521},
  year     = {2021},
  doi      = {10.1021/acs.jpcb.1c04019}
}

@article{RN45,
  author   = {Shirts, M. R. and Chodera, J. D.},
  title    = {Statistically optimal analysis of samples from multiple equilibrium states},
  journal  = {J. Chem. Phys.},
  volume   = {129},
  number   = {12},
  year     = {2008},
  doi      = {10.1063/1.2978177}
}

@article{RN139,
  author   = {Thompson, A. P. and Aktulga, H. M. and Berger, R. and Bolintineanu, D. S. and Brown, W. M. and Crozier, P. S. and Veld, P. J. I. and Kohlmeyer, A. and Moore, S. G. and Nguyen, T. D. and Shan, R. and Stevens, M. J. and Tranchida, J. and Trott, C. and Plimpton, S. J.},
  title    = {LAMMPS-a flexible simulation tool for particle-based materials modeling at the atomic, meso, and continuum scales},
  journal  = {Comput. Phys. Commun.},
  volume   = {271},
  year     = {2022},
  doi      = {10.1016/j.cpc.2021.108171}
}

@article{RN122,
  author   = {Yagasaki, T. and Matsumoto, M. and Tanaka, H.},
  title    = {Lennard-Jones Parameters Determined to Reproduce the Solubility of NaCl and KCl in SPC/E, TIP3P, and TIP4P/2005 Water},
  journal  = {J. Chem. Theory Comput.},
  volume   = {16},
  number   = {4},
  pages    = {2460--2473},
  year     = {2020},
  doi      = {10.1021/acs.jctc.9b00941}
}

@article{RN15,
  author   = {Young, J. M. and Panagiotopoulos, A. Z.},
  title    = {System-Size Dependence of Electrolyte Activity Coefficients in Molecular Simulations},
  journal  = {J. Phys. Chem. B},
  volume   = {122},
  number   = {13},
  pages    = {3330--3338},
  year     = {2018},
  doi      = {10.1021/acs.jpcb.7b09861}
}

@article{RN110,
  author   = {Zwanzig, R. W.},
  title    = {High-temperature equation of state by a perturbation method. {I}. Nonpolar gases},
  journal  = {J. Chem. Phys.},
  volume   = {22},
  number   = {8},
  pages    = {1420--1426},
  year     = {1954},
  doi      = {10.1063/1.1740409}
}

\end{document}